\documentclass[useAMS,usenatbib,twocolumn]{mn2e}
\pdfoutput=1
\usepackage{amssymb}
\usepackage{amsmath}
\usepackage{natbib}
\usepackage[pdftex]{graphicx}
\usepackage{subfigure}
\usepackage{booktabs}
\usepackage{tabularx}
\begin{document}

\title[Eccentric discs]
{Nonlinear hydrodynamical evolution of eccentric Keplerian discs in two dimensions: validation of secular theory}
      \author[A. J. Barker \& G. I. Ogilvie]{A. J. Barker\thanks{Email address: ajb268@cam.ac.uk} and G. I. Ogilvie\\
Department of Applied Mathematics and Theoretical Physics, University of Cambridge, Centre for Mathematical Sciences, \\ Wilberforce Road, Cambridge CB3 0WA, UK}
	
\date{}
\pagerange{\pageref{firstpage}--\pageref{lastpage}} \pubyear{2016}

\maketitle

\label{firstpage}

\begin{abstract}
We perform global two-dimensional hydrodynamical simulations of Keplerian discs with free eccentricity over thousands of orbital periods. Our aim is to determine the validity of secular theory in describing the evolution of eccentric discs, and to explore their nonlinear evolution for moderate eccentricities. Linear secular theory is found to correctly predict the structure and precession rates of discs with small eccentricities. However, discs with larger eccentricities (and eccentricity gradients) are observed to precess faster (retrograde relative to the orbital motion), at a rate that depends on their eccentricities (and eccentricity gradients). We derive analytically a nonlinear secular theory for eccentric gas discs, which explains this result as a modification of the pressure forces whenever eccentric orbits in a disc nearly intersect. This effect could be particularly important for highly eccentric discs produced in tidal disruption events, or for narrow gaseous rings;  
it might also play a role in causing some of the variability in superhump binary systems.
In two dimensions, the eccentricity of a moderately eccentric disc is long-lived and persists throughout the duration of our simulations. Eccentric modes are however weakly damped by their interaction with non-axisymmetric spiral density waves (driven by the Papaloizou-Pringle instability, which occurs in our idealised setup with solid walls), as well as numerical diffusion. 
\end{abstract}


\section{Introduction}
Eccentric gas discs are thought to arise in a number of astrophysical contexts. The orbital evolution of a newly born planet due to its tidal interaction with the protoplanetary disc is intricately coupled with the evolution of eccentric motions within the disc. But the importance of this interaction in producing the eccentricities of observed exoplanets is currently unclear \citep{PapNM2001,Pap2002,GoldreichSari2003,KleyDirksen2006,DAngelo2006,Bitsch2013}. Eccentric gas discs are also thought to explain the superhump phenomenon in SU UMa-type binary stars \citep{Whitehurst1988,Lubow1991a,Goodchild2006,Smith2007} and the spectral variability of rapidly rotating Be stars \citep{Okazaki1991,PapSav1992,Ogilvie2008}. In addition, eccentric gas discs are formed by the tidal disruption of stars or gaseous planets \citep{Guillochon2011,Liu2013,GuillochonMR2014}, and such a process might be responsible for the gas cloud G2 near Sgr A* in the Galactic Centre \citep{Guillochon2014,CoughlinNixon2015,Pfuhl2015}. Despite this wide range of applications, the nonlinear dynamics of eccentric discs remains poorly understood.

In the Solar system, several of the rings of Saturn and Uranus are observed to be elliptical, and a theory for narrow eccentric rings (with small eccentricities but arbitrary eccentricity gradients) composed of weakly collisional particles has been developed \citep{BGT1983,ChiangGoldreich2000}. For small eccentricities, eccentric discs can be described as slowly precessing one-armed density waves (or shocks) (e.g.~\citealt{Okazaki1991,PapSav1992,LeeGoodman1999,Goodchild2006,Ogilvie2008,Saini2009}). A particular solution for a global uniformly eccentric (and apsidally aligned) Keplerian disc has also been studied by \cite{Statler2001}.

A general secular theory for three-dimensional eccentric discs with arbitrary eccentricities and eccentricity gradients has been developed \citep{Ogilvie2001}, and we aim to explore (and test using numerical simulations) some of its implications in this work. Secular theory describes dynamical processes that occur on timescales that are much longer than the orbital timescale. In the more familiar context of a system of multiple planets around a star, the Keplerian orbit of each planet is subject to perturbing forces that are proportional to the planet-to-star mass ratio; secular theory accounts for the long-term behaviour of eccentricity and inclination resulting from the orbit-averaged perturbing forces. For a thin gaseous disc around a star, however, pressure provides the most important departure from Keplerian orbital motion, being of second order in the aspect ratio $H/R$ of the disc. Other subdominant forces, such as those due to the self-gravity of the disc \citep{Tremaine2001} and viscous or turbulent stresses \citep{Ogilvie2001}, can also be important. In this situation, secular theory describes the long-term evolution of eccentricity due to gas pressure (and other forces) in thin discs, but neglects contributions that are of fourth order or higher in $H/R$.

In recent work, we developed a local model of an eccentric disc \citep{OB2014a}, which is similar to the shearing box that is commonly used to study the dynamics of circular astrophysical discs \citep{GoldreichLyndenBell1965,Hawley1995}. We used this model to analyse the vertical oscillatory flows that are driven by the variation in the vertical gravity around a non-circular orbit \citep{OB2014a}, and subsequently studied the local linear stability of these discs and their vertical flows \citep{BO2014a}. We found that eccentric discs are generically unstable (in three dimensions), being subject to a small-scale instability in which inertial waves are driven by a parametric resonance. This instability is expected to lead to wave activity or turbulence, and to damping of the disc eccentricity and eccentricity gradients \citep{John2005b}, unless these are maintained by external forcing (or additional instabilities e.g. viscous overstability). Nonlinear simulations are required to determine the efficiency of this damping process so that its importance in real discs can be quantified (e.g.~continuing the work begun by \citealt{John2005b}, by taking into account the vertical structure of the disc).

In this paper we study the nonlinear hydrodynamical evolution of eccentric discs in two dimensions, deferring three-dimensional simulations to future work.  Our aim is to determine the validity of linear secular theory \citep{Ogilvie2001,Pap2002,OB2014a} in describing the structure and precession rates of eccentric discs, as well as to determine their two-dimensional nonlinear dynamics. In the process, we will derive analytically, and compare with simulations, a two-dimensional nonlinear secular theory for isothermal eccentric discs that is valid for any eccentricity and eccentricity gradient, for a thin untwisted disc with non-intersecting orbits -- this is a particular case of the general theory of \cite{Ogilvie2001} that is amenable to analytical study. In two dimensions, eccentric discs do not exhibit the parametric instability or vertical oscillatory flows. However, a fundamental study of the two-dimensional nonlinear dynamics of eccentric discs has not yet been undertaken. We believe this to be worthwhile, in spite of the clear importance of three-dimensional effects in reality, because the two-dimensional dynamics that we will study are likely to also play a role in three dimensions. In particular, the force due to gas pressure, which tends to cause retrograde precession of these discs, can be captured in two dimensions. These simulations also provide a necessary benchmark to allow comparison with future three-dimensional simulations.

We outline our intentionally simplified numerical setup designed to study the dynamics of eccentric discs in \S~\ref{Numerical}. We derive analytically a nonlinear secular theory for untwisted eccentric discs in two-dimensions in Appendix \ref{nonlineartheory}. This theory is explored, and its predictions for the shapes and precession rates of eccentric discs are compared with simulations, in \S~\ref{Secular}. The long-term nonlinear evolution of eccentric discs is presented in \S~\ref{Results}, where we also analyse the background instability that arises due to our idealised setup with rigid walls in the absence of any free eccentricity. We then finish with our conclusions.

\section{Simplified model}
\label{Numerical}

\subsection{Background disc}
Our model consists of an eccentric (nearly) Keplerian disc in a two-dimensional cylindrical domain as an initial condition. We study its resulting nonlinear evolution using the PLUTO code \citep{Mignone2007}, adopting cylindrical polar coordinates ($R,\phi$) with corresponding velocity components $u_R$ and $u_\phi$. Our governing equations are
\begin{eqnarray}
&& \left(\partial_t + \boldsymbol{u}\cdot\nabla\right)\boldsymbol{u} = -\frac{1}{\Sigma}\nabla p-\frac{GM}{R^2}\boldsymbol{e}_R, \\
&& \partial_t \Sigma + \nabla \cdot (\Sigma\boldsymbol{u})=0.
\end{eqnarray}
To allow us to more easily understand the outcome of our simulations, we consider a power-law disc which is isothermal (where the pressure $p=c_{s}^{2}\Sigma$, and $\Sigma$ is the total surface density; with sound speed $c_{s}=$ const), with background disc surface density,
\begin{eqnarray}
\label{sigma}
\Sigma_b(R)&=&\Sigma_{0}R^{-\sigma},
\end{eqnarray}
for the circular case, where $\sigma$ will be varied. (We expect the qualitative results of this paper to carry over to discs with different thermodynamic behaviour, and those with a radially varying sound speed.) The axisymmetric basic state of the disc has angular velocity $\Omega=\Omega_0R^{-\frac{3}{2}}$ (which does not strictly apply when $\sigma\ne0$ due to radial pressure gradients), where $\Omega_0$ is the Keplerian angular velocity at $R=R_i$, and $u_\phi=R\Omega$. Our disc occupies the full azimuthal extent\footnote{Meaning that periodic boundary conditions are applied to all quantities at $\phi=0$ and $\phi=2\pi$.} $\phi\in[0,2\pi)$, with radial extent $R\in [R_i,R_o]$, on which the radial boundary conditions are ``reflecting" conditions\footnote{This means that the radial velocity in the ghost cells adjacent to the domain has the same magnitude but the opposite sign to that in the last grid cell inside the domain. The surface density in the ghost cells is set to be the same as that in the last grid cell, i.e.~ $\partial_R\Sigma= 0$ at this location. This may lead to minor differences with secular theory, where no boundary condition is imposed on $\Sigma$.}. These boundary conditions were chosen to confine the eccentric disc for a well-defined study. While these are not appropriate for all applications in which eccentric discs are thought to arise, they may reasonably approximate the boundaries of a disc that has been tidally truncated (e.g.~\citealt{John2005b}). For example, our model might be relevant to narrow rings that may be produced by disc-planet interactions that have formed multiple gaps. A reflecting edge may also be appropriate to describe the inner edge of the disc, if there is a sharp drop in the density there, or where discs match onto the surfaces of central objects e.g. white dwarfs in Cataclysmic Variable systems. However, the aim of this paper is not to focus on any particular application, but to explore the fundamental dynamics of eccentric discs that might have more general applicability. We choose units such that $\Omega_0=R_i=\Sigma_0=1$, and vary the parameters $c_s$ (to mimic a disc with aspect ratio $H/R\in[0.025,0.1]$), $\sigma$ and $R_o$, in addition to the eccentricity of the disc, which we will now describe.

\subsection{Eccentric mode}
We initialise an eccentric disc using nonlinear secular theory (Appendix \ref{nonlineartheory}). 
To do this we modify the velocity components and surface density of the background disc so that they describe an eccentric mode, to make the streamlines elliptical. The eccentric mode is a global slowly precessing density wave that satisfies the boundary conditions. 

We define the complex eccentricity $E(\lambda)=e (\lambda) \mathrm{e}^{\mathrm{i} \omega(\lambda)}$, where $e$ and $\omega$ are the eccentricity and longitude of pericentre. In secular theory, orbits are labelled using their semi-latus rectum $\lambda$, related to cylindrical radius by
\begin{eqnarray}
R(\lambda,\phi) = \frac{\lambda}{1+ e(\lambda) \cos \left(\phi -\omega(\lambda)\right)}.
\end{eqnarray}
In linear secular theory for a 2D isothermal disc, 
\begin{eqnarray}
\label{linearsecular}
2 \Sigma \left(G M \lambda^3 \right)^{\frac{1}{2}} \partial_t  E = \mathrm{i} c_s^2 \left(\partial_\lambda \left(\Sigma \lambda^3 E'\right) + \lambda^2 E \partial_{\lambda}\Sigma \right),
\end{eqnarray}
where $E'\equiv \partial_{\lambda} E$, and in this case $R$ and $\lambda$ are equivalent (this can be transformed into the Schr\"{o}dinger equation). Eq.~\ref{nonlinearsecular} is the equivalent equation in the nonlinear secular theory of Appendix \ref{nonlineartheory} (in which care must be taken to take into account the difference between $R$ and $\lambda$, and the resulting equation is a type of nonlinear Schr\"{o}dinger equation). We seek solutions that precess at the rate $\omega_p$, such that $E\propto \mathrm{e}^{\mathrm{i}\omega_p t}$. Together with the boundary conditions
\begin{eqnarray}
E(\lambda_i)=E(\lambda_o)=0,
\end{eqnarray}
appropriate for fixed circular boundaries, we obtain an eigenvalue problem for the eigenvalue $\omega_p$ and eigenfunction $E$. We solve this problem using a shooting method \citep{Press1992}, which works equally well for the solution of Eq.~\ref{nonlinearsecular} (which is a nonlinear equation for $E$ and $E'$). This method requires us to provide an initial guess for $\omega_p$ and $E'(\lambda_i)$ to select the appropriate solution. To do this, we use a combination of trial and error and the results of an independently coded solution to the linear problem using a Chebyshev collocation method. In each case, we select the fundamental eccentric mode, which has a single maximum in the eccentricity. This is the slowest precessing mode with the longest radial wavelength, which is chosen because it can be simulated more accurately than any mode of shorter wavelength. For the linear theory, we normalise the amplitude of the resulting mode so that its maximum eccentricity is $A$. In the nonlinear secular theory, its amplitude is determined by $E'(\lambda_i)$, which we choose so that the maximum eccentricity is $A$ (to within an accuracy of approximately $10^{-4}$). 

A nonlinear eccentric mode cannot be exactly represented as a single Fourier mode with an azimuthal wavenumber $m=1$. However, we can obtain $u_R,u_\phi,\Sigma$ for the eccentric Keplerian disc as follows. First, we obtain the velocity components of the eccentric orbital motion at each grid point $(R,\phi)$ using $E(\lambda)$:
\begin{eqnarray}
\label{Kepler1}
u_R &=&\sqrt{\frac{GM}{\lambda}} e(\lambda) \sin (\phi-\omega(\lambda)), \\
u_\phi &=&\sqrt{\frac{GM}{\lambda}} \left(1+e(\lambda) \cos (\phi-\omega(\lambda))\right),
\label{Kepler2}
\end{eqnarray}
after we have converted points in $\lambda,\phi$ to $R,\phi$ \citep{MurrayDermott1999}. We obtain the surface density by considering mass conservation \citep{OB2014a}, by noting that $\mathcal{M}/\mathcal{P}=J\Omega\Sigma$ (see Appendix \ref{nonlineartheory} for definitions). We assume that the mass distribution in the disc is a power law in $\lambda$ of the form $\mathcal{M}(\lambda)=2\pi \lambda \Sigma_b(\lambda)$, so that 
\begin{eqnarray}
\label{Kepler3}
\Sigma = \Sigma_b(\lambda) \frac{\left(1-e(\lambda)^2\right)^{\frac{3}{2}}\left(1+e(\lambda) \cos (\phi-\omega(\lambda)\right)}{1+\left(e(\lambda)-\lambda \partial_{\lambda}e(\lambda) \right)\cos(\phi-\omega(\lambda)},
\end{eqnarray}
which is the natural extension of our cylindrical disc model.
We use Eqs.~\ref{Kepler1}--\ref{Kepler3} as our initial conditions in PLUTO, noting that there are errors $O(c_s^2)$ because we have made the secular (thin-disc) approximation \citep{Ogilvie2001}. The eccentric disc is exactly represented in nonlinear secular theory for any eccentricity and eccentricity gradient. Note that we neglect the radial pressure gradient which would slightly modify the angular velocity profile when $\sigma\ne 0$.

We use a second-order Runge-Kutta time-stepping algorithm, with linear interpolation within grid-cells. We adopted the dimensionally-unsplit HLLC solver (which was chosen because preliminary investigation found it to be more robust than the Roe solver, and also much less diffusive than the HLL solver). The eccentric mode is input into PLUTO after appropriate interpolation to the grid used in the code.

\section{Predictions and validation of nonlinear secular theory}
\label{Secular}

In this section, we will illustrate the predictions of secular theory and compare them with the results of hydrodynamical simulations. We will focus only on aspects that are directly relevant for such a comparison, deferring a more detailed discussion of the simulations to \S~\ref{Results}.

\subsection{Modification of pressure forces in nonlinear theory}
\label{Modification}

\begin{figure}
  \begin{center}
    \subfigure{\includegraphics[trim=2cm 0cm 3cm 0cm, clip=true,width=0.35\textwidth]{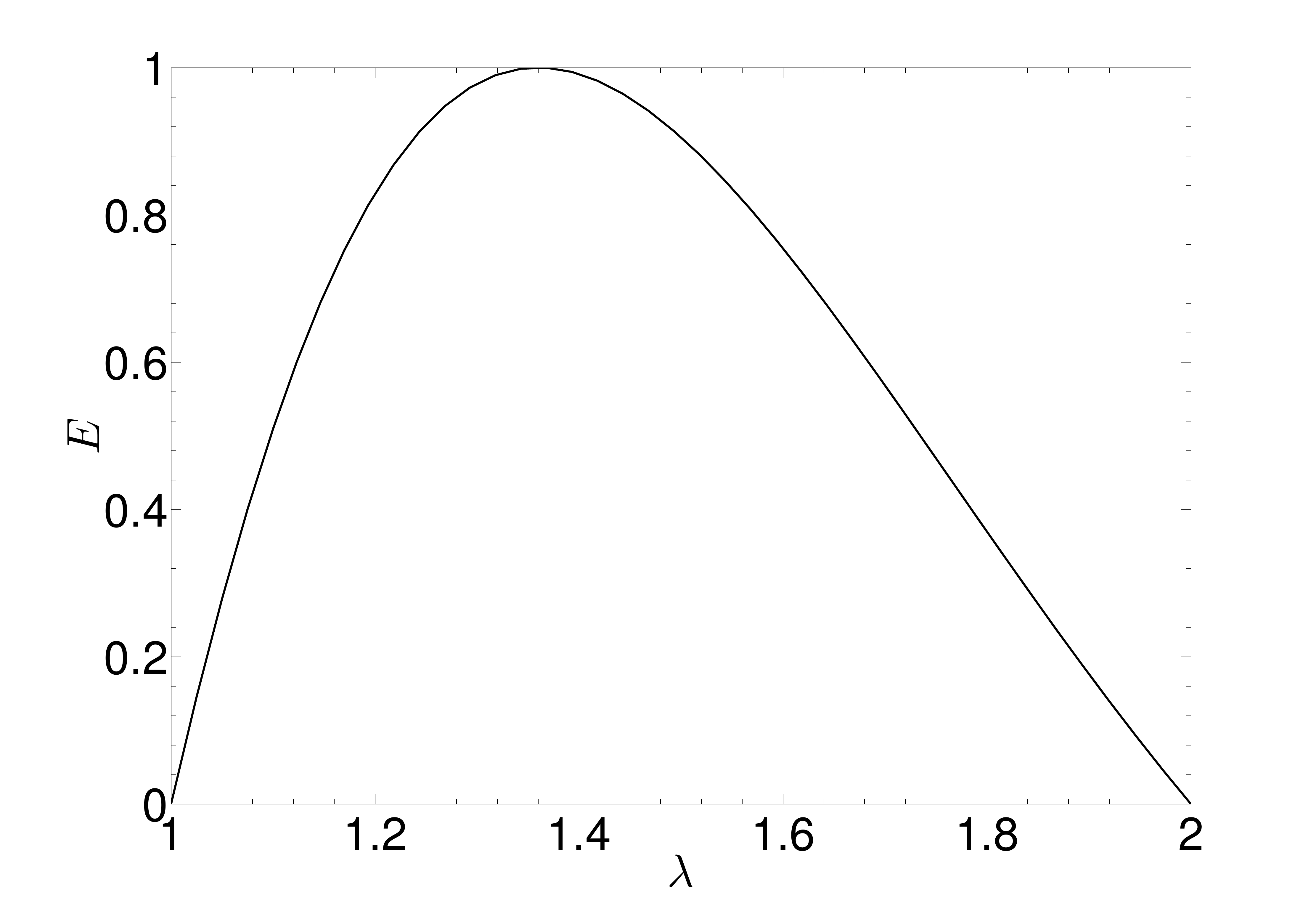}}
    \subfigure{\includegraphics[trim=1cm 0cm 3cm 0cm, clip=true,width=0.36\textwidth]{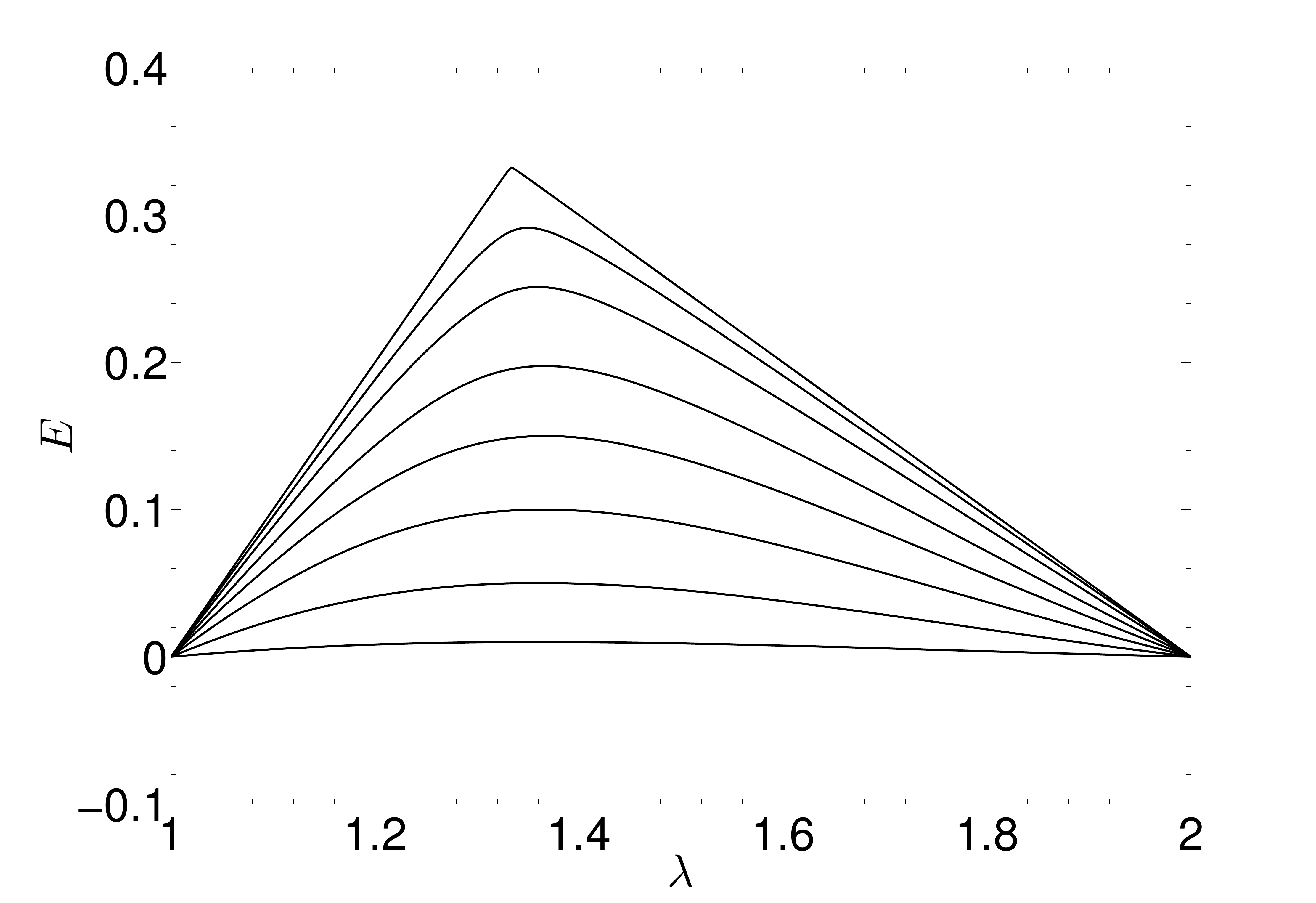} }
     \subfigure{\includegraphics[trim=7cm 0cm 8cm 0cm, clip=true,width=0.35\textwidth]{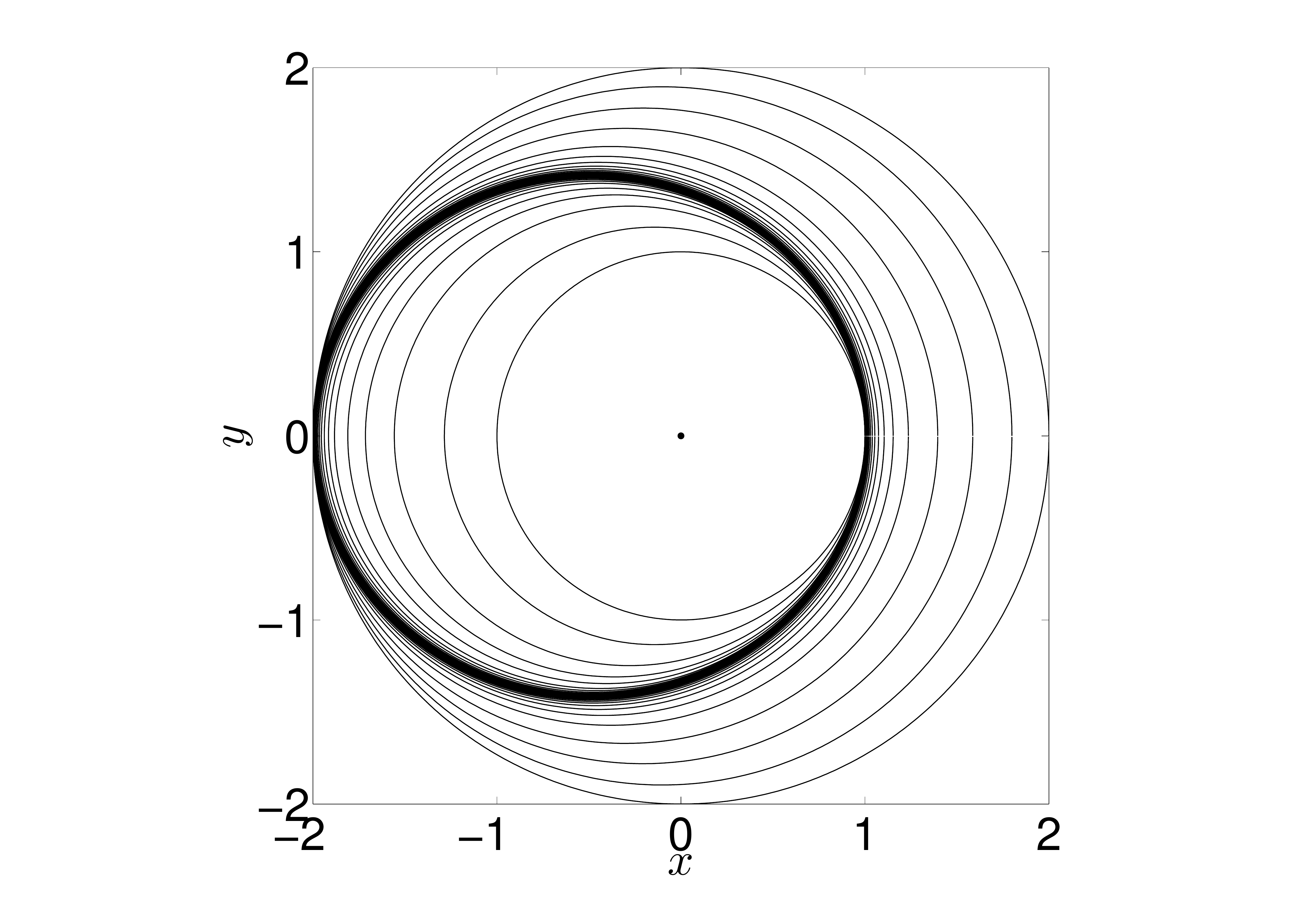} }
    \end{center}
  \caption{Illustrative fundamental eccentric mode solutions $e(\lambda)$ in a domain with $R_o=2$ and $\sigma=0$ (and $c_s=0.05$) according to linear secular theory (top) and nonlinear secular theory (middle). The amplitude in the top panel is arbitrary. Middle: solutions with $A\approx 0.01,0.05,0.1,0.15,0.2,0.25,0.3$ and $0.332$ have been plotted. Bottom: orbital geometry in the $x,y$-plane for the eccentric mode with $A\approx0.332$, showing orbits spaced equally in $\lambda$, which clearly shows the compression and near-intersection of orbits in this case.}
  \label{secularfig}
\end{figure}

Linear secular theory (Eq.~\ref{linearsecular}, together with boundary conditions) predicts the structure and precession rate of eccentric modes as a function of the disc properties. For the case of an untwisted disc with $R_o=\lambda_o=2$ and $\sigma=0$, we have plotted the fundamental eccentric mode $E(\lambda)$ in the top panel of Fig.~\ref{secularfig}. The amplitude is arbitrary, so this represents the shape of the disc for any amplitude according to linear theory. However, neighbouring orbits (for an untwisted disc) intersect if 
\begin{eqnarray}
\label{intersect}
\left(e-\lambda e'\right)^2 \geq 1,
\end{eqnarray}
which is observed to occur for this linear mode at $R=R_i$ if $A\gtrsim 0.143$ (for $R_o=2$; similarly this occurs when $A\gtrsim 0.18$ for $R_{0}=3$, and $A\gtrsim 0.24$ for $R_{0}=10$). This suggests that secular theory will no longer apply to these discs, since orbital intersections will rapidly lead to eccentricity damping via shocks. However, this prediction is based on linear theory. 

Nonlinear secular theory (Eq.~\ref{nonlinearsecular}, together with the boundary conditions) predicts the shape of the eccentric mode to depend on its amplitude $A$, as we illustrate in the middle panel of Fig.~\ref{secularfig} for several amplitudes as indicated in the caption. Inspection of the functional form of the terms on the right hand side of Eq.~\ref{nonlinearsecular} informs us why this is the case: these terms increasingly differ from those in linear theory as the orbits approach an intersection, i.e.~the shape depends on the amplitude because pressure forces act to minimise intersections. This also causes the precession of the mode to differ from the predictions of linear theory, as we will demonstrate in \S~\ref{Precession}.

With our adopted boundary conditions, nonlinear theory predicts a maximum attainable amplitude for the eccentricity, corresponding to the upper curve in the middle panel of Fig.~\ref{secularfig}, which occurs when $A\approx0.332$. We have plotted the orbital geometry for the eccentric disc with $A\approx0.332$, with orbits spaced equally in $\lambda$, in the bottom panel of Fig.~\ref{1}. This shows clearly the compression and near-intersection of orbits near the inner and outer boundaries for this mode, in addition to the fact that it can no longer be described purely as an $m=1$ mode. The maximum value can be explained simply by considering the (untwisted) eccentric mode which is marginally intersecting, so that $e -\lambda e^{\prime}=\pm 1$, that also satisfies the boundary conditions that $e(\lambda_i)=e(\lambda_o)=0$. The solution is the piecewise linear profile
\begin{eqnarray}
e = \begin{cases} 
   \frac{\lambda}{\lambda_{i}}-1, & \text{if } \lambda \in [\lambda_i,\lambda_m], \\
   1-\frac{\lambda}{\lambda_{o}}, & \text{if } \lambda\in(\lambda_m,\lambda_{o}],
  \end{cases}
\end{eqnarray}
where
\begin{eqnarray}
\lambda_m=\frac{2}{\lambda_i^{-1}+\lambda_o^{-1}}.
\end{eqnarray}
The maximum eccentricity of this mode is
\begin{eqnarray}
e_\mathrm{max}\equiv e(\lambda_m)=\frac{\lambda_o-\lambda_i}{\lambda_i+\lambda_o},
\end{eqnarray}
which explains our observation of a maximum eccentricity for the fundamental eccentric mode in a domain with a given size (in numerical calculations we observe $e_\mathrm{max}(\lambda_o=2)\approx 0.332$, $e_\mathrm{max}(\lambda_o=3)\approx 0.496$ and $e_\mathrm{max}(\lambda_o=5)\approx 0.66$). This solution is what we must obtain from geometrical considerations simply because we force the eccentricity to vanish at the boundaries. However, this result depends on the boundary conditions, since this limit would no longer apply if $e'(\lambda_o)=0$, for example, and we would not obtain a maximum eccentricity smaller than one.

\subsection{Precession rates: validation of theory}
\label{Precession}

According to linear theory, the eccentric mode precesses slowly in a retrograde sense due to gas pressure. To see that this must be the case, we can multiply Eq.~\ref{linearsecular} by $E^{*}$, seek solutions $E\propto \mathrm{e}^{\mathrm{i}\omega_p t}$ and integrate over $\lambda$ to obtain:
\begin{eqnarray}
\omega_p=\frac{c_s^2\int_{\lambda_i}^{\lambda_o} \left(\lambda^2 |E|^2 \partial_\lambda\Sigma - \lambda^3\Sigma |E'|^2 \right) \mathrm{d}\, \lambda}{2\int_{\lambda_i}^{\lambda_o} \left(GM\lambda^3\right)^{\frac{1}{2}}\Sigma |E|^2\mathrm{d}\, \lambda},
\end{eqnarray}
with our chosen boundary conditions \citep{Goodchild2006}. This quantity is a negative real number if $\partial_\lambda\Sigma\leq 0$, i.e.~$\sigma\geq 0$, indicating retrograde precession\footnote{If we were to model a real system by including additional effects that make the potential non-Keplerian, such as rotational deformation of the star \citep{PapSav1992}, presence of planets within the disc \citep{Pap2002} or the inclusion of self-gravity \citep{Tremaine2001}, in addition to three-dimensional gas pressure effects associated with vertical oscillatory flows \citep{Ogilvie2008}, the precession rate would be altered and could even become prograde.}.

\begin{figure}
  \begin{center}
    \subfigure{\includegraphics[trim=3cm 0cm 6cm 0cm, clip=true,width=0.45\textwidth]{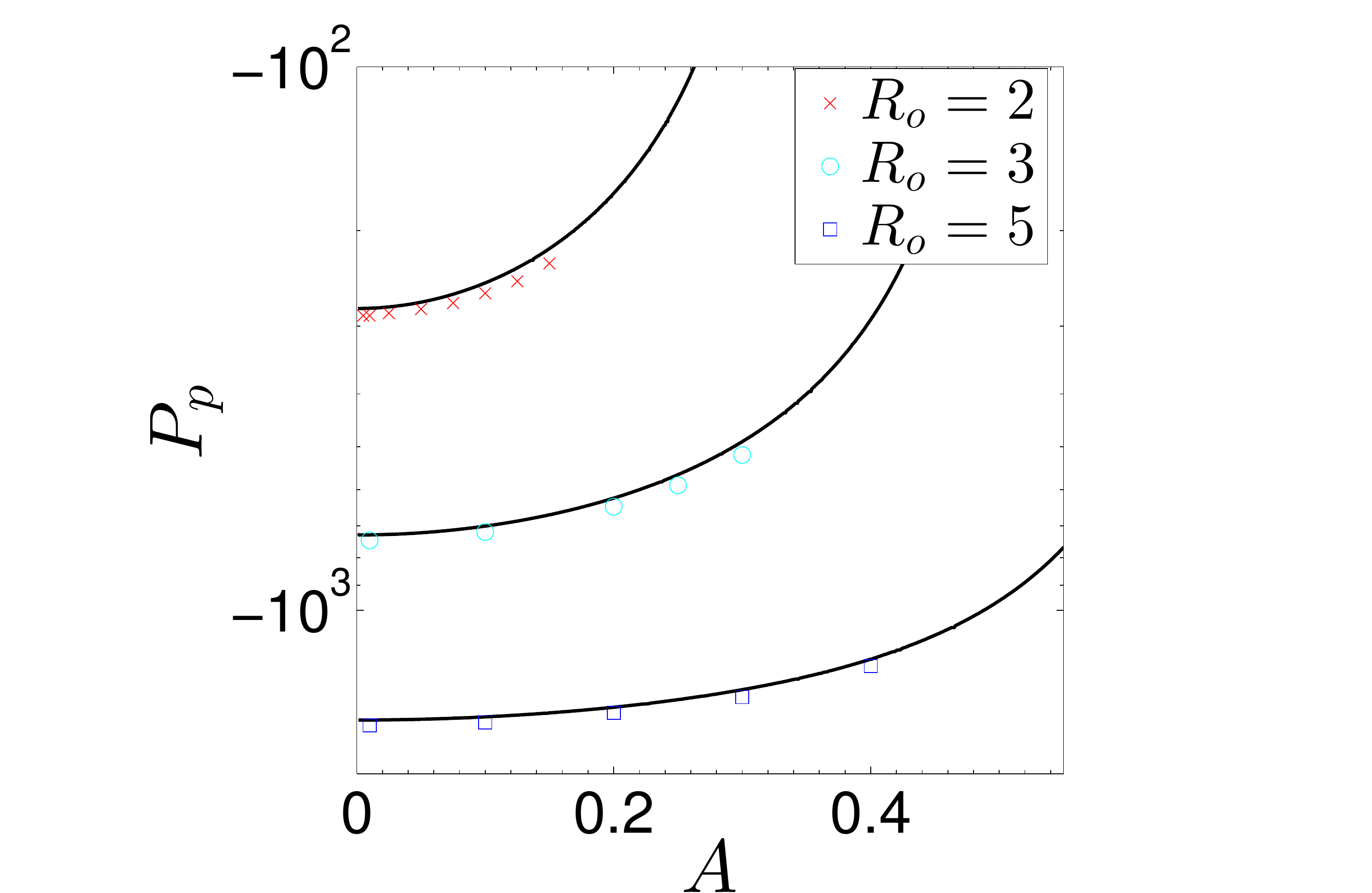} }
   \end{center}
  \caption{Precession period $P_p$ versus peak eccentric mode amplitude $A$ showing a comparison between nonlinear secular theory (black solid lines) and the results of hydrodynamical simulations in a domain $R_o=2,3$ and 5 (red crosses, light blue circles and blue squares, respectively), $c_s=0.05$ and $\sigma=0$. This demonstrates that the eccentric mode precesses faster as its amplitude is increased. We observe good agreement between simulations and nonlinear secular theory as $A$ and $R_o$ are varied (with a departure of approximately $3\%$). Note that secular theory predicts its own breakdown for large amplitudes, as $P_p\rightarrow 0$.}
  \label{nonlinearprecession}
\end{figure}

\begin{table*}
\begin{tabular}{ccccc|cccl}
 $c_{s}$ & $R_{o}$ & $\sigma$ & $A$ & $N_R\times N_\phi$ & $P^{\mathrm{obs}}_{p} (\pm 1)$ & $P^{\mathrm{pred}}_{p}$ & $P^{\mathrm{NS,Lin,pred}}_{p}$ \\
  \hline
 0.05 & $2$ & 0 & $0.005$ & 200x256 & 287 & 278.33 & 286.01 \\
 0.05 & $2$ & 0 & $0.01$ & 200x256 & 287 & 278.11 & \\ 
 0.05 & $2$ & 0 & $0.025$ & 200x256 & 284 & 276.56 & \\
 0.05 & $2$ & 0 & $0.05$ & 200x256 & 280 & 271.03 & \\
 0.05 & $2$ & 0 & $0.075$ & 200x256 & 272 & 262.07 &  \\
 0.05 & $2$ & 0 & $0.1$ & 200x256 & 261 & 249.78 & \\
 0.05 & $2$ & 0 & $0.125$ & 200x256 & 248 & 234.35 & \\
 0.05 & $2$ & 0 & $0.15$ & 200x256 & 230 & 216.00 & \\
 \hline
 0.05 & $3$ & 0 & $0.01$ & 300x256 & 744 & 726.78 & 742.22 \\
 0.05 & $3$ & 0 & $0.1$ & 300x256 & 718 & 700.20 &  \\
 0.05 & $3$ & 0 & $0.2$ & 300x256 & 645 & 621.45 & \\
 0.05 & $3$ & 0 & $0.25$ & 300x256 & 589 & 562.74 & \\
 0.05 & $3$ & 0 & $0.3$ & 300x256 & 518 & 490.66 & \\
 \hline
 0.05 & $5$ & 0 & $0.01$ & 500x256 & 1629 & 1592.91 & 1627.26 \\
 0.05 & $5$ & 0 & $0.1$ & 500x256 & 1608 & 1571.43 & \\
 0.05 & $5$ & 0 & $0.2$ & 500x256 & 1546 & 1509.13 & \\
 0.05 & $5$ & 0 & $0.3$ & 500x256 & 1446 & 1402.52 & \\
 0.05 & $5$ & 0 & $0.4$ & 500x256 & 1267 & 1231.76 & \\
 \hline
 0.025 & $2$ & 0 & $0.01$ & 200x256 & 1123 & 1112.4 & 1121.47 \\ 
 0.075 & $2$ & 0 & $0.01$ & 200x256 & 133 & 143.39 & 130.99 \\ 
 0.1 & $2$ & 0 & $0.01$ & 200x256 & 77 & 69.53 & 76.44 \\ 
 0.05 & $2$ & 1.5 & $0.01$ & 200x256 & 277 & 271.05 & 279.46 \\
 0.05 & $2$ & 3 & $0.01$ & 200x256 & 257 & 251.87 & 260.58 \\
 0.05 & $2$ & 5 & $0.01$ & 200x256 & 224 & 215.68 & 224.52 \\
 \hline
 0.05 & $2$ & 0 & $0.05$ & 800x1024 & 280 & 271.03 & \\
 0.05 & $2$ & 0 & $0.05$ & 400x512 & 280 & 271.03 & \\
 0.05 & $2$ & 0 & $0.05$ & 100x128 & 280 & 271.03 & \\
 0.05 & $2$ & 0 & $0.05$ & 50x64 & 287 & 271.03 & \\
\end{tabular}
\caption{Table of simulations used to validate nonlinear secular theory. In each case the precession is retrograde. The error bars for $P_p^\mathrm{obs}$ correspond with the time interval between output files used for this analysis. $P_p^\mathrm{pred}$ is the prediction from nonlinear secular theory (Eq.~\ref{nonlinearsecular}) and $P^{\mathrm{NS,Lin,pred}}_{p}$ is the (non-secular) prediction from solution of the two-dimensional linearised isothermal hydrodynamic equations in Appendix~\ref{nonseculartheory} (we have entered the latter in the smallest $A$ entry in the table for each case, but it should be remembered that these results are based on a linear calculation). Simulations were run for at least one full precession period.}
\label{table1}
\end{table*}

The precession rate depends on $\sigma$, $R_o$ and $c_s$, but is independent of the amplitude of the eccentric mode in linear theory. However, we have shown that nonlinear secular theory predicts the structure of the eccentric mode to be modified if the orbits are close to intersecting, so we might expect $\omega_p$ to depend on its amplitude also. We demonstrate that this is indeed the case in the black solid lines in Fig.~\ref{nonlinearprecession}, where we have plotted the precession period $P_p=\frac{2\pi}{\omega_p}$ against the amplitude of the eccentric mode for several calculations with different $R_o\in [2,3,5]$ and various $A$ with $\sigma=0$ and assuming a thin disc with $c_s=0.05$. Gas pressure causes the mode to precess faster, in a retrograde sense, as the eccentricity and eccentricity gradients are increased. We obtain the linear prediction in the limit $A\rightarrow 0$, with a departure that is initially $\propto A^2$, before steepening for large $A$. The precession period continues to shorten until $A\rightarrow e_\mathrm{max}$, when $P_p\rightarrow 0$, so that the secular approximation is no longer valid. Nonlinear secular theory therefore predicts its own breakdown for large amplitudes, which occurs when the orbits in the disc come close to intersecting. 

In Fig.~\ref{nonlinearprecession} we have also plotted the initial precession period computed from hydrodynamical simulations (symbols according to the legend) that were run for at least one full precession period. A table of simulations used for this comparison, and those in the rest of this section, is given in Table~\ref{table1}. We calculate the precession period by computing the $m=1$ component of the radial velocity (this quantity is used to represent the power in both $m=\pm1$), 
\begin{eqnarray}
\hat{u}_{1}(R,t)=\frac{1}{\pi}\int_0^{2\pi}u_{R}(R,\phi,t)\mathrm{e}^{-\mathrm{i} \phi} \mathrm{d}\,\phi,
\end{eqnarray}
from which the mean longitude of pericentre can be computed from
\begin{eqnarray}
\langle \omega(t)\rangle =\frac{1}{R_o-R_i} \int_{R_i}^{R_o}\tan^{-1} \left(-\frac{\mathrm{Im}[\hat{u}_{1}(R,t)]}{\mathrm{Re}[\hat{u}_{1}(R,t)]}\right)\,\mathrm{d}\, R,
\end{eqnarray}
which is approximately valid for the eccentric mode even when $A$ is not small. Since this mode precesses retrogradely, $\langle\omega\rangle$ decreases cyclically from $\pi$ to $-\pi$, and we calculate the initial precession period by eye by checking at what time $\langle \omega \rangle \approx \pi$ after $t=0$. The data used for this analysis is output at each time unit in the simulation, giving errors of $\pm1$ in the determination of $P_\mathrm{obs}$, which we further verified by visual inspection of $u_R(R,\phi,t)$ at these time snapshots. Simulations with $A\gtrsim0.15$ for $R_o=2$ were not plotted, since the eccentric mode underwent non-negligible damping during a single precession period, as we will discuss further in \S~\ref{Results}. Fig.~\ref{nonlinearprecession} shows that our results are in good agreement with the nonlinear secular theory for each $R_o$ considered.

To further validate the predictions of secular theory, we have listed the observed and predicted values of $P_p$ as various disc (and simulation) parameters are varied in Table~\ref{table1}. The variation of $P_p$ with $\sigma$ and $c_s$ is reasonably captured in each case. However, the departure from the predictions of secular theory increases as we increase $c_s$: for $c_s=[0.025,0.05,0.075,0.1]$, the fractional error correspondingly takes the approximate values $[1,3.2,7.5,10]\%$, indicating a roughly $c_s^2$ dependence for this quantity, as expected from secular theory \citep{Ogilvie2001,OB2014a}. To confirm that this departure indeed results from the partial inaccuracy of the secular approximation, we have also computed the precession frequency of an $m=1$ linear eccentric mode by solving the two-dimensional linearised isothermal (non-secular) hydrodynamic equations as outlined in Appendix~\ref{nonseculartheory}, for the cases listed in the table. This accurately matches the observed precession rates of our smallest $A$ simulations in each case. For thin discs with $c_s\lesssim 0.1$, the precession period predicted by secular theory matches the observed values to within a few percent, typically tending to predict slightly slower precession.

We have also computed the precession rate for a set of calculations with $A=0.05$ as the resolution is varied, as indicated in Table~\ref{table1}. This indicates convergence is attained for the precession period even at low resolutions such as $100\times 128$, but a departure appears for even lower resolutions. This suggests that simulations with typically-adopted resolutions would be expected to capture the precession rates of global eccentric modes, but may not accurately capture the precession of shorter wavelength modes. 

\subsection{Summary}

We have shown that the retrograde precession of eccentric discs as a result of their gas pressure is enhanced for sufficiently large eccentricities and eccentricity gradients. This dependence of the precession rate on the eccentricity and its gradient is a new result\footnote{We also note that a precession rate that depends on the eccentricity gradient can be derived using the formalism of \cite{BGT1983} if this is applied to an isothermal gas.} (though there was some numerical evidence of this in \citealt{John2005b}), and occurs because pressure forces are enhanced when the orbits come close to intersecting. While the numerical results would change if we were to adopt different boundary conditions, this general behaviour is likely to be robust since it arises from the functional form of the stress integrals (see Appendix~\ref{nonlineartheory}).

We have thoroughly validated the predictions of nonlinear secular theory regarding the precession rates of eccentric discs as various parameters are varied, in the regime where $\omega_p\ll \Omega$. We find that the departure from secular theory scales as $O(c_s^2)$, typically being a few percent or smaller for thin discs if $\omega_p\ll \Omega$ (which we have shown to break down at large amplitudes). We now turn to a more detailed analysis of our simulation results, focussing on the long-term evolution of the eccentricity in two dimensions.

\section{More detailed discussion of simulation results}
\label{Results}

\subsection{Background instability: nonlinear evolution of the ``Papaloizou-Pringle" instability}
\label{PPinst}

Our basic setup has rigid walls to confine an eccentric mode and permit a well-defined study. However, it is well known that a circular supersonic rotating shear flow in a container with one or more rigid boundaries is unstable to the development of non-axisymmetric instabilities \citep{PP1984,PP21985,GoldreichNarayan1985,GGN1986,PP31987,Kato1987,NGG1987}. These instabilities are driven either by Kelvin-Helmholtz-type mechanisms (which require a potential vorticity minimum, which is not the case in our problem), or by over-reflection of spiral density waves from the corotation region. While most of the original work on this problem focused on thick discs (referred to as ``accretion tori") where this instability excites low azimuthal wavenumber ($m=O(1)$) modes on a local orbital timescale, slower growing instabilities with high azimuthal wavenumbers ($m\gg 1$) have also been found to occur in thin Keplerian discs \citep{PP31987,NGG1987,Hanawa1987}. The nonlinear evolution of these instabilities has been studied by \cite{Godon1998} (and earlier for thick ``accretion tori" by \citealt{Hawley1987}), where they were observed to drive subsonic wave activity in hydrodynamic discs.

The basic mechanism of instability in a thin nearly Keplerian disc can be explained as follows. An outgoing spiral density wave with frequency $\omega_m$ and azimuthal wavenumber $m\ne 0$ with a radial angular momentum flux $\mathcal{I}$ that approaches its corotation region (the region where the wave frequency satisfies $\omega_m\approx m\Omega$) is primarily reflected from its inner Lindblad resonance (where $\omega_m-m\Omega = -\kappa$, and $\kappa\approx \Omega$ is the epicyclic frequency), with the angular momentum flux in the reflected wave being $\mathcal{R}$. However, this wave is also partially transmitted through the corotation region (in which it is evanescent), requiring an outgoing wave to be launched at the outer Lindblad resonance (where $\omega_m-m\Omega=\kappa$) with transmitted angular momentum flux $\mathcal{T}=\mathcal{I}-\mathcal{R}$. In a Keplerian disc, waves propagating inside corotation ($\omega_m\leq m\Omega$) carry a negative angular momentum flux (so that $\mathcal{I}<0$), whereas those outside corotation ($\omega_m\geq m\Omega$) carry a positive flux (so that $\mathcal{T}>0$). This means that $\mathcal{R}=\mathcal{I}-\mathcal{T}<\mathcal{I}$, so that the reflected wave is amplified. If there is an inner impermeable boundary, the reflected wave will be completely reflected once more so that it will re-enter the corotation region, enabling further amplification at the expense of the Keplerian flow. This mechanism of instability, which is due to the ``over-reflection" of waves from the corotation region, also occurs if there is a rigid outer boundary, but does not occur if both boundaries perfectly transmit wave energy, and it can be eliminated with sufficient viscosity.

\begin{figure}
  \begin{center}
     \subfigure[L]{\includegraphics[trim=6.5cm 0cm 7cm 0cm, clip=true,width=0.35\textwidth]{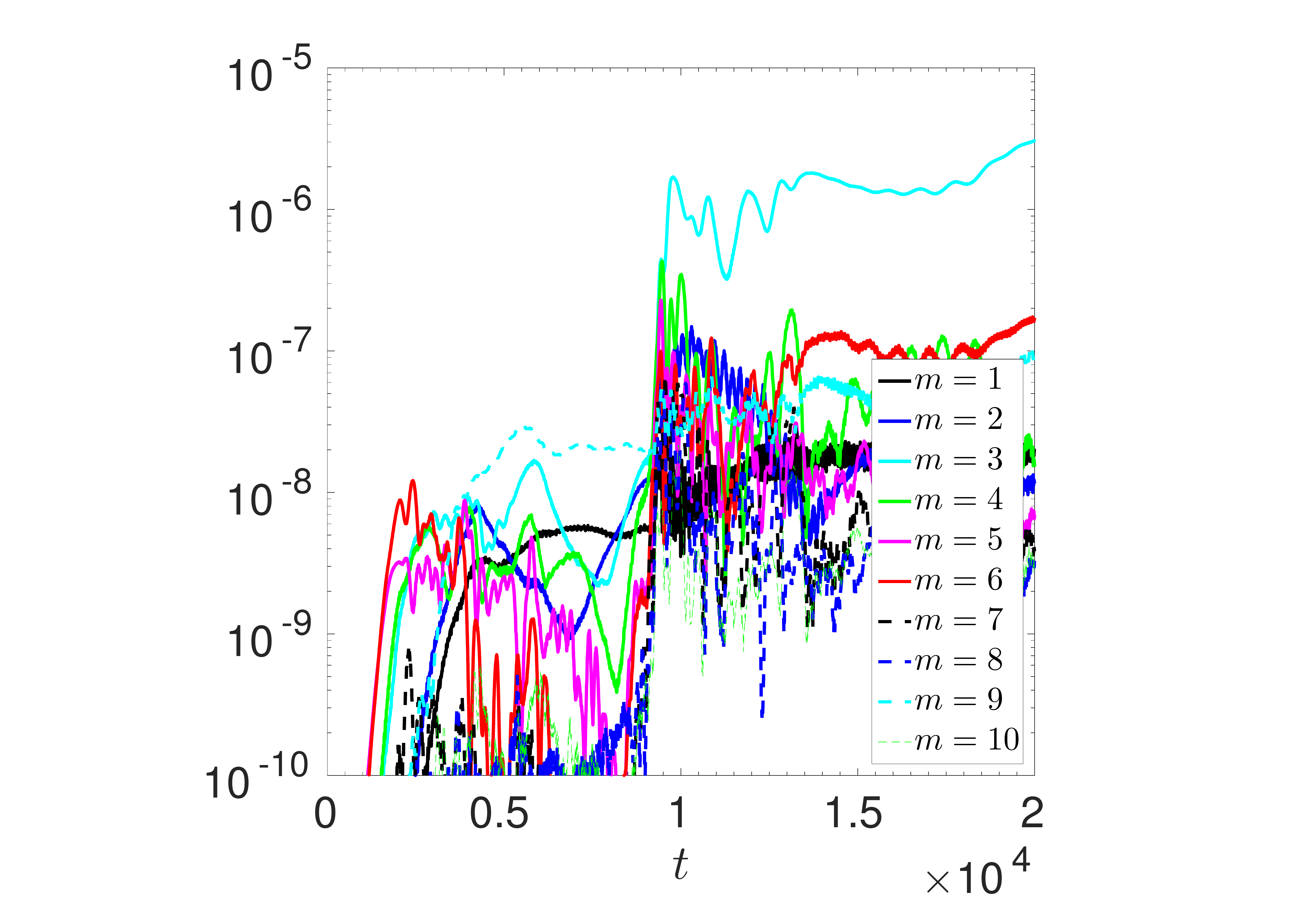}}
     \subfigure[M]{\includegraphics[trim=6.5cm 0cm 7cm 0cm, clip=true,width=0.35\textwidth]{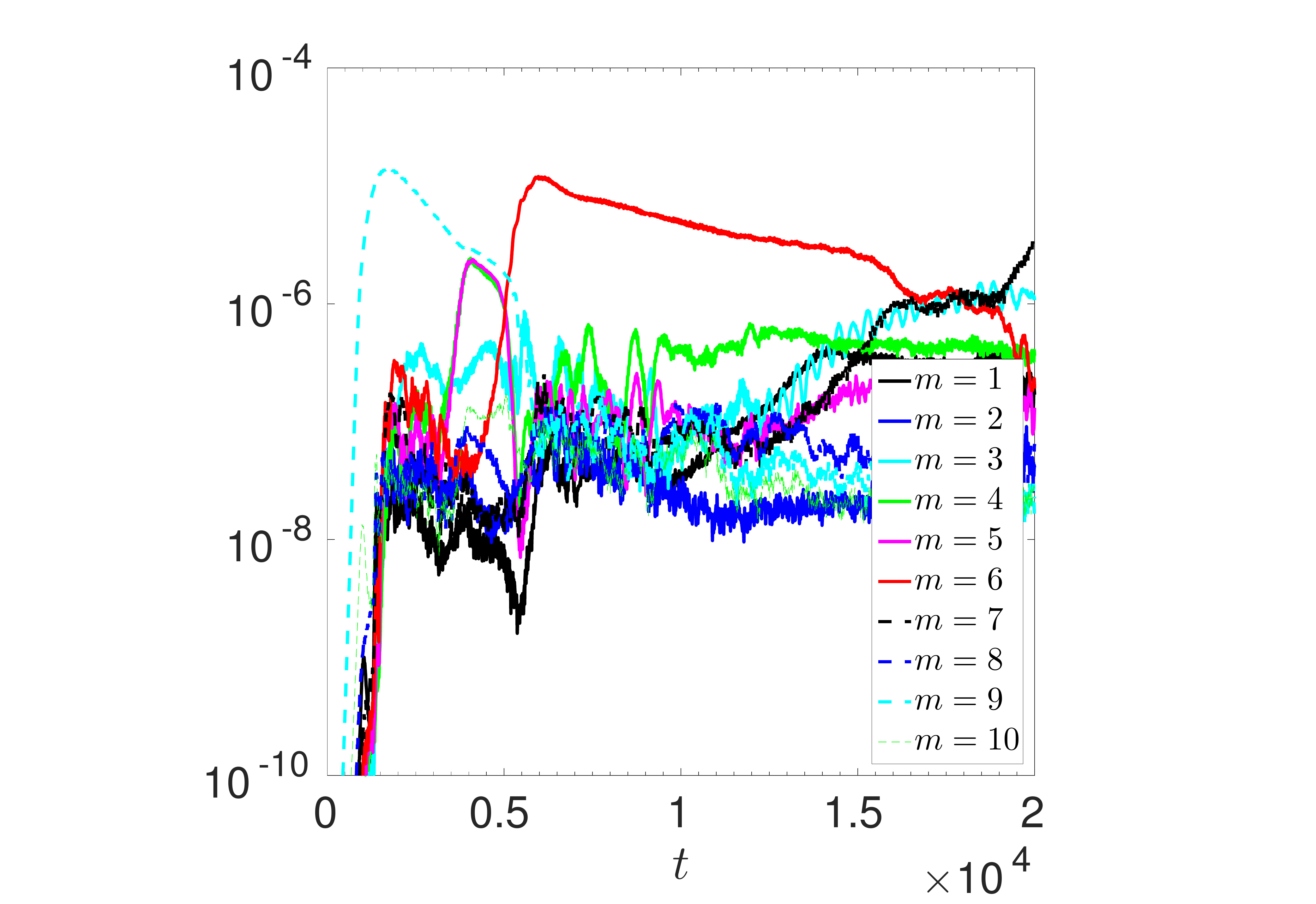}} 
     \subfigure[H]{\includegraphics[trim=6.5cm 0cm 7cm 0cm, clip=true,width=0.35\textwidth]{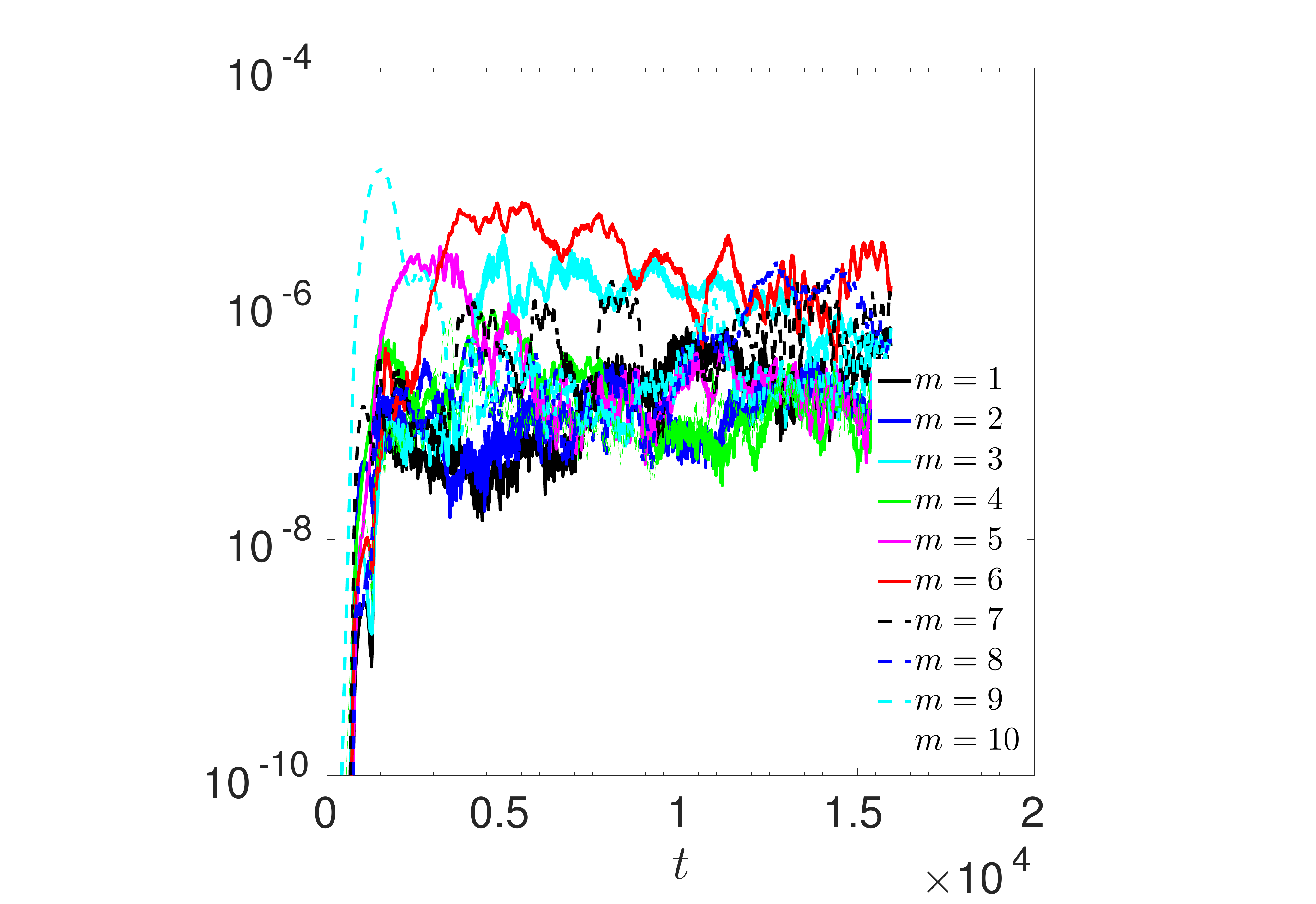}}
    \end{center}
  \caption{Energy in different azimuthal wavenumber components of the solution ($E_m$) for $m\in [1,10]$ as a function of time for simulations in a domain with $R_o=2$ with $c_s=0.05$ and $\sigma=0$ for three different resolutions $200\times 256$ (L), $400\times 512$ (M) and $800\times 1024$ (H).}
  \label{PP1}
\end{figure}

\begin{figure}
  \begin{center}
   \subfigure[L, $R_o=3$]{\includegraphics[trim=6.5cm 0cm 7cm 0cm, clip=true,width=0.35\textwidth]{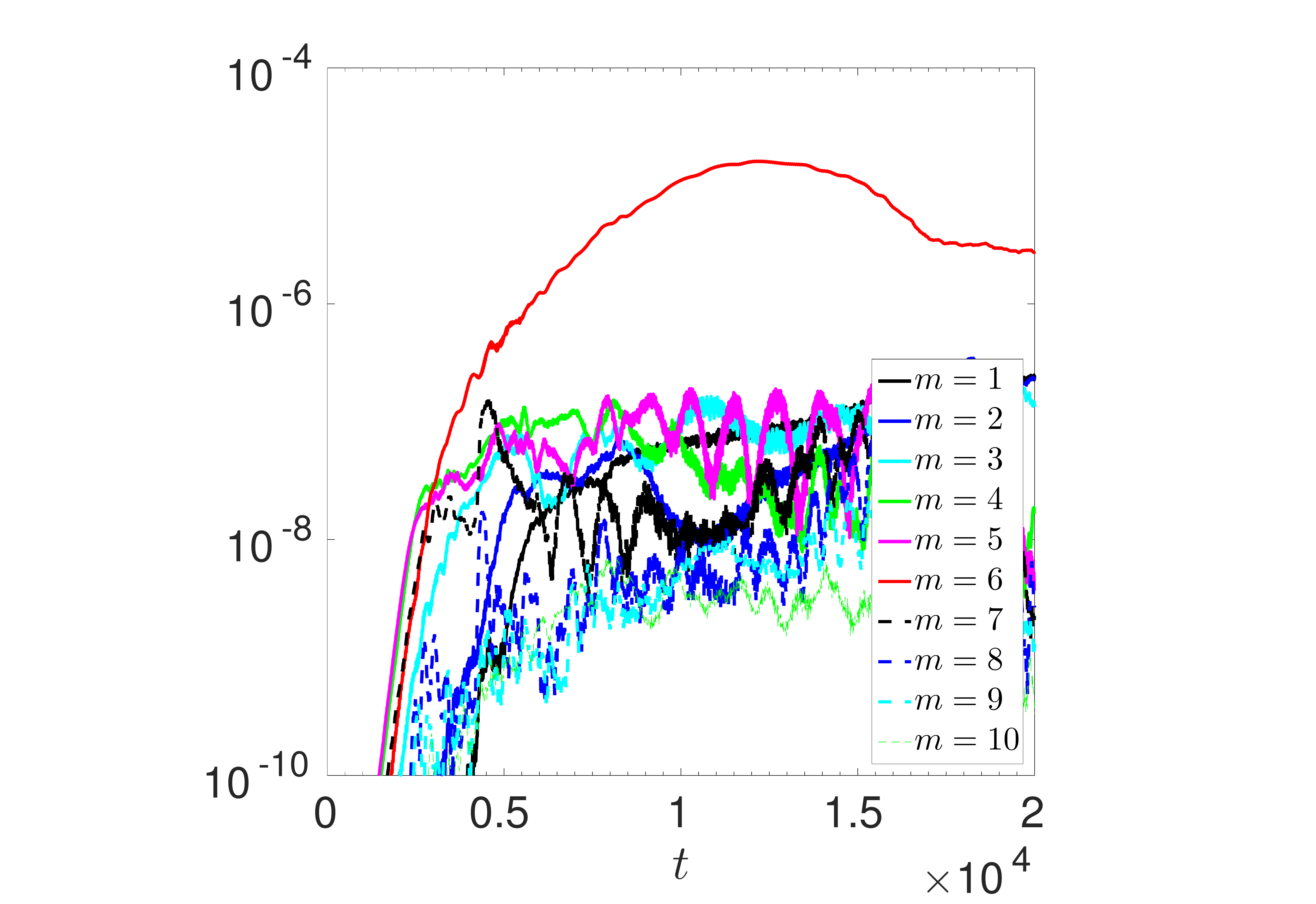}}
     \subfigure[M, $R_o=3$]{\includegraphics[trim=6.5cm 0cm 7cm 0cm, clip=true,width=0.35\textwidth]{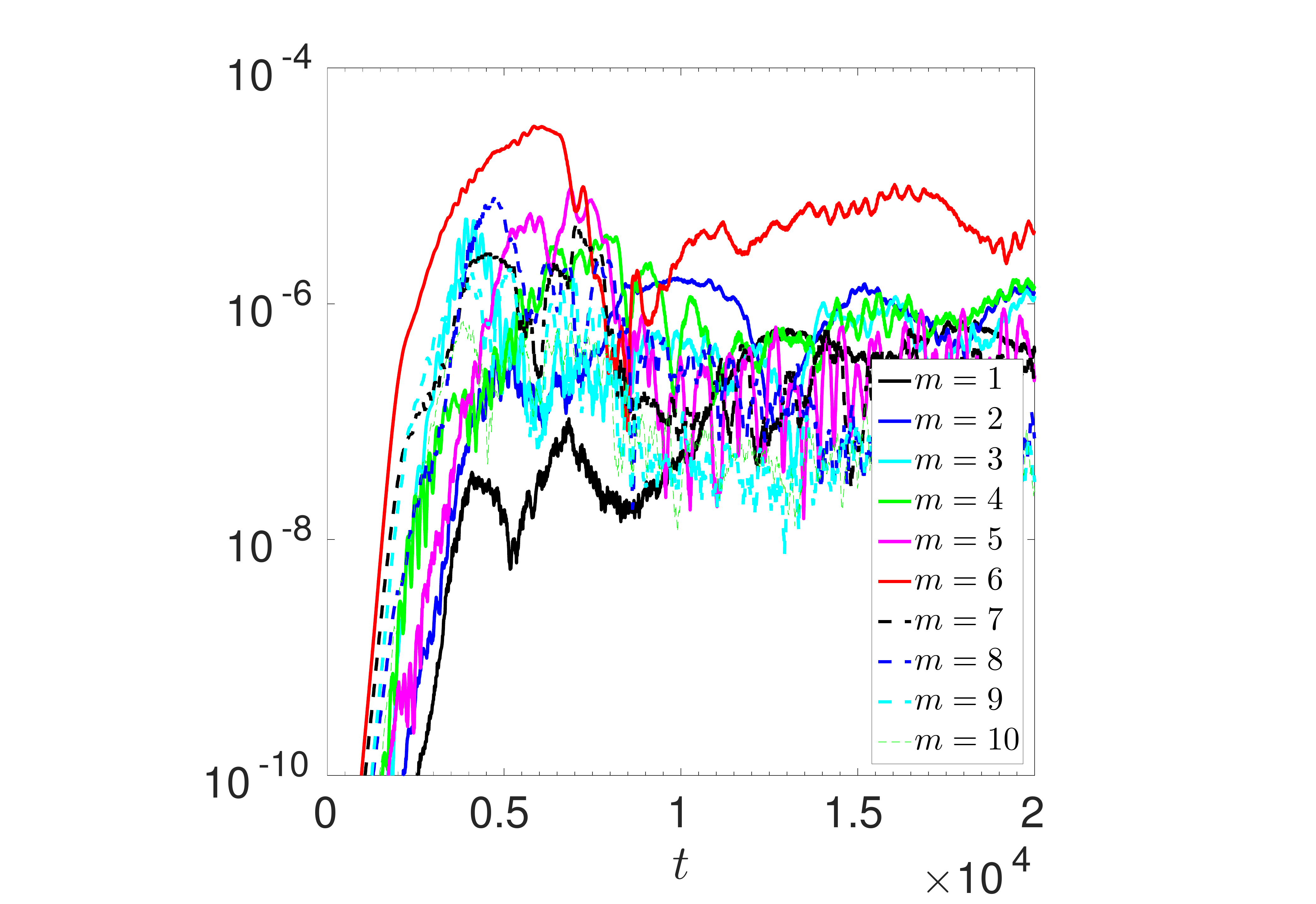}}
    \end{center}
  \caption{Same as Fig.~\ref{PP1}, but for simulations with $R_o=3$ with resolutions of $300\times 256$ (L) and $600\times 512$ (M).}
  \label{PP1a}
\end{figure}

\begin{figure}
  \begin{center}
     \subfigure{\includegraphics[trim=0cm 0cm 0cm 0cm, clip=true,width=0.35\textwidth]{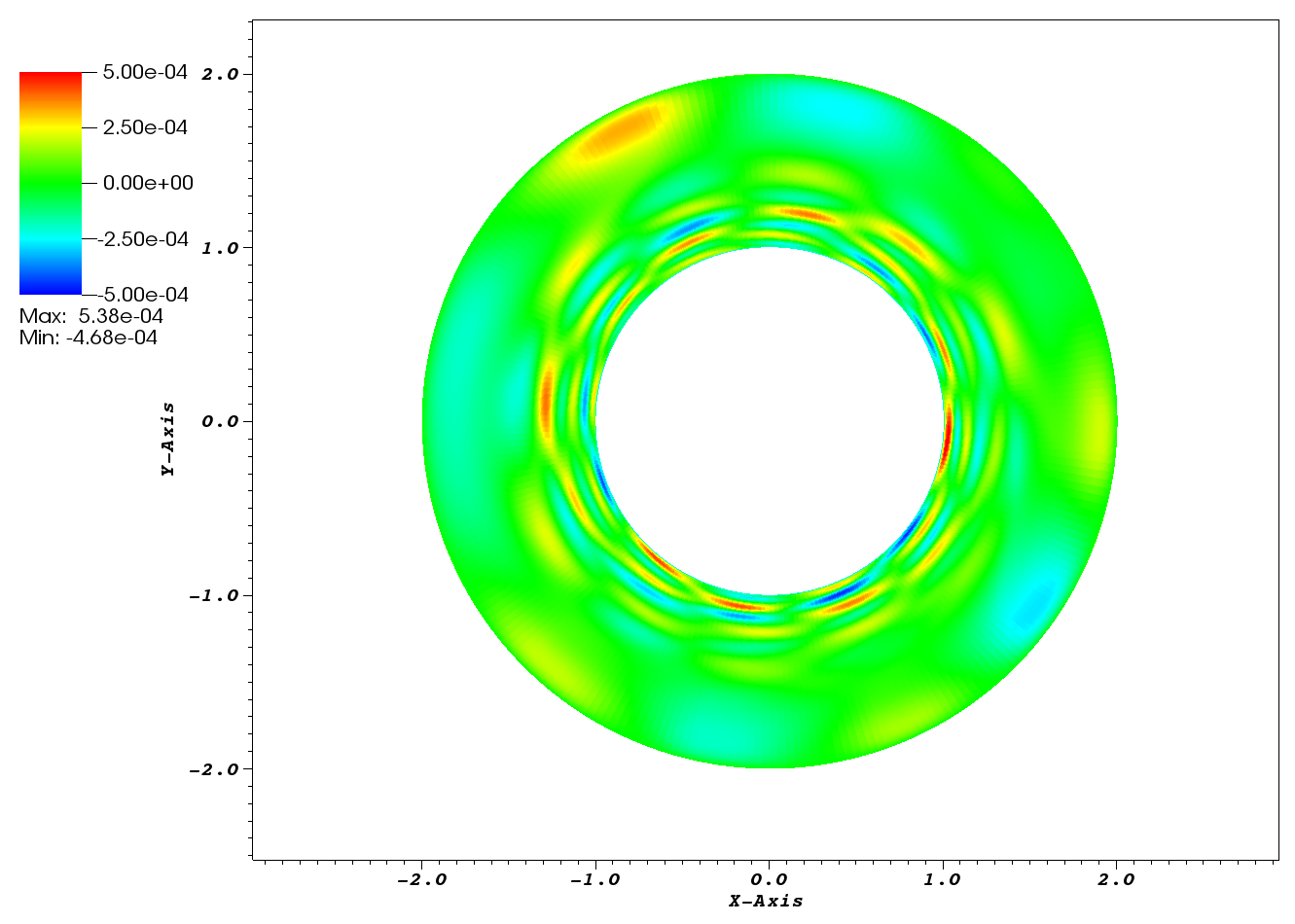}}
     \subfigure{\includegraphics[trim=0cm 0cm 0cm 0cm, clip=true,width=0.35\textwidth]{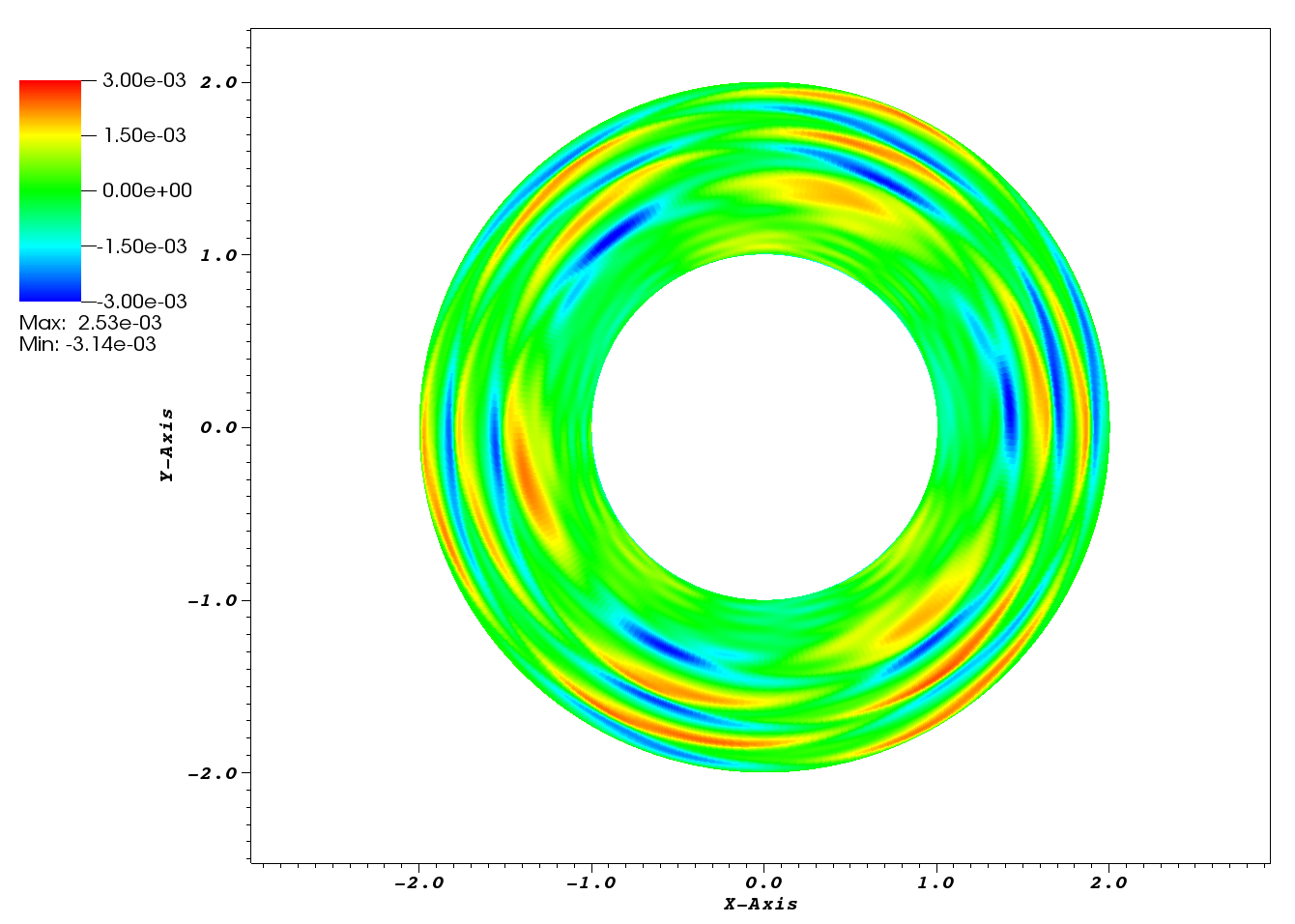}}
         \end{center}
  \caption{Radial velocity $u_R$ in the linear growth phase (top; propagating spiral waves are produced at inner boundary) and at a later stage (bottom; global radial standing waves are shown) of the instability in our lowest resolution simulation, illustrating the non-axisymmetric waves generated by the instability.}
  \label{PP3}
\end{figure}

\begin{figure}
  \begin{center}
     \subfigure{\includegraphics[trim=5.5cm 0cm 7cm 0cm, clip=true,width=0.35\textwidth]{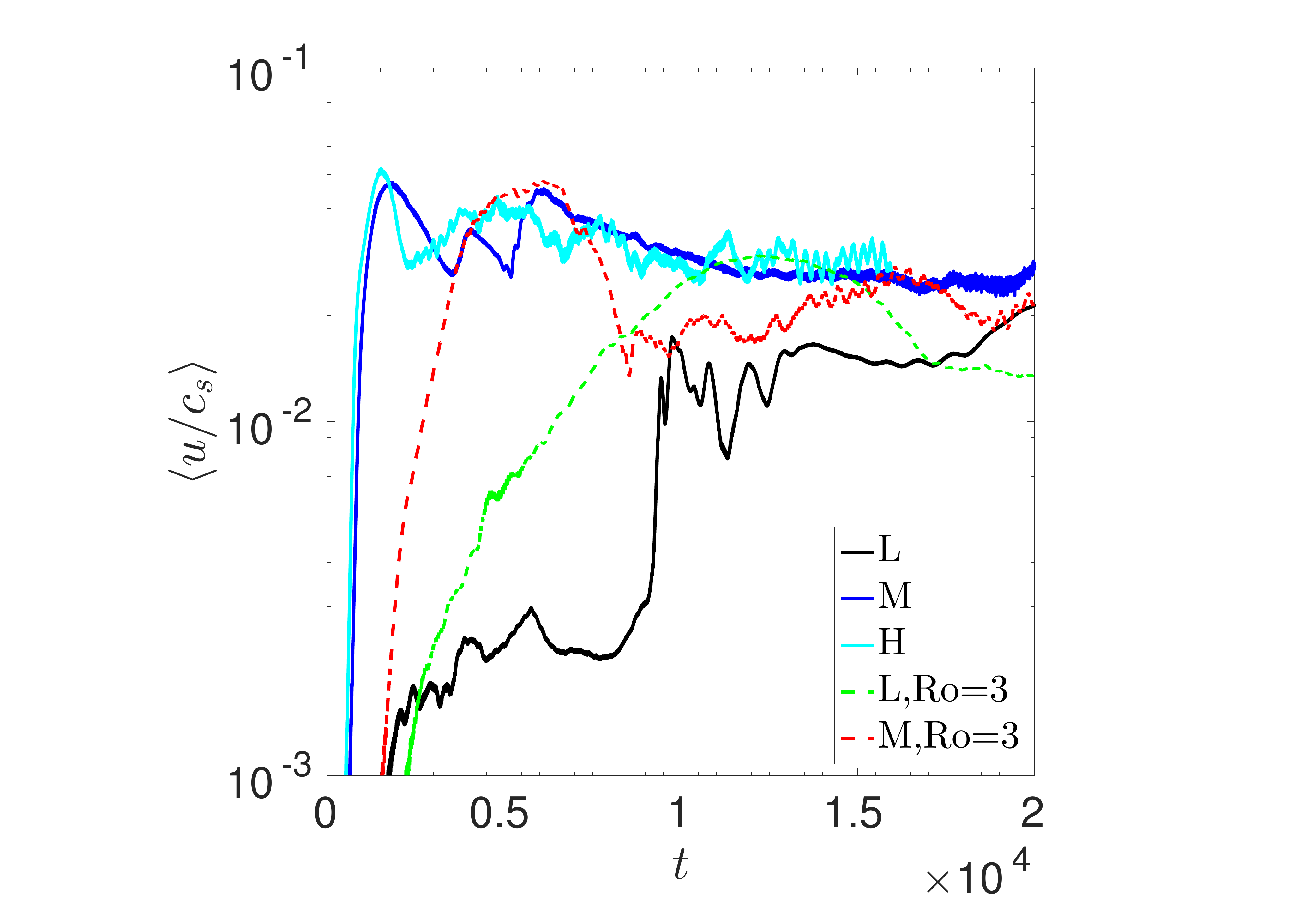}}
     \subfigure{\includegraphics[trim=5.5cm 0cm 7cm 0cm, clip=true,width=0.35\textwidth]{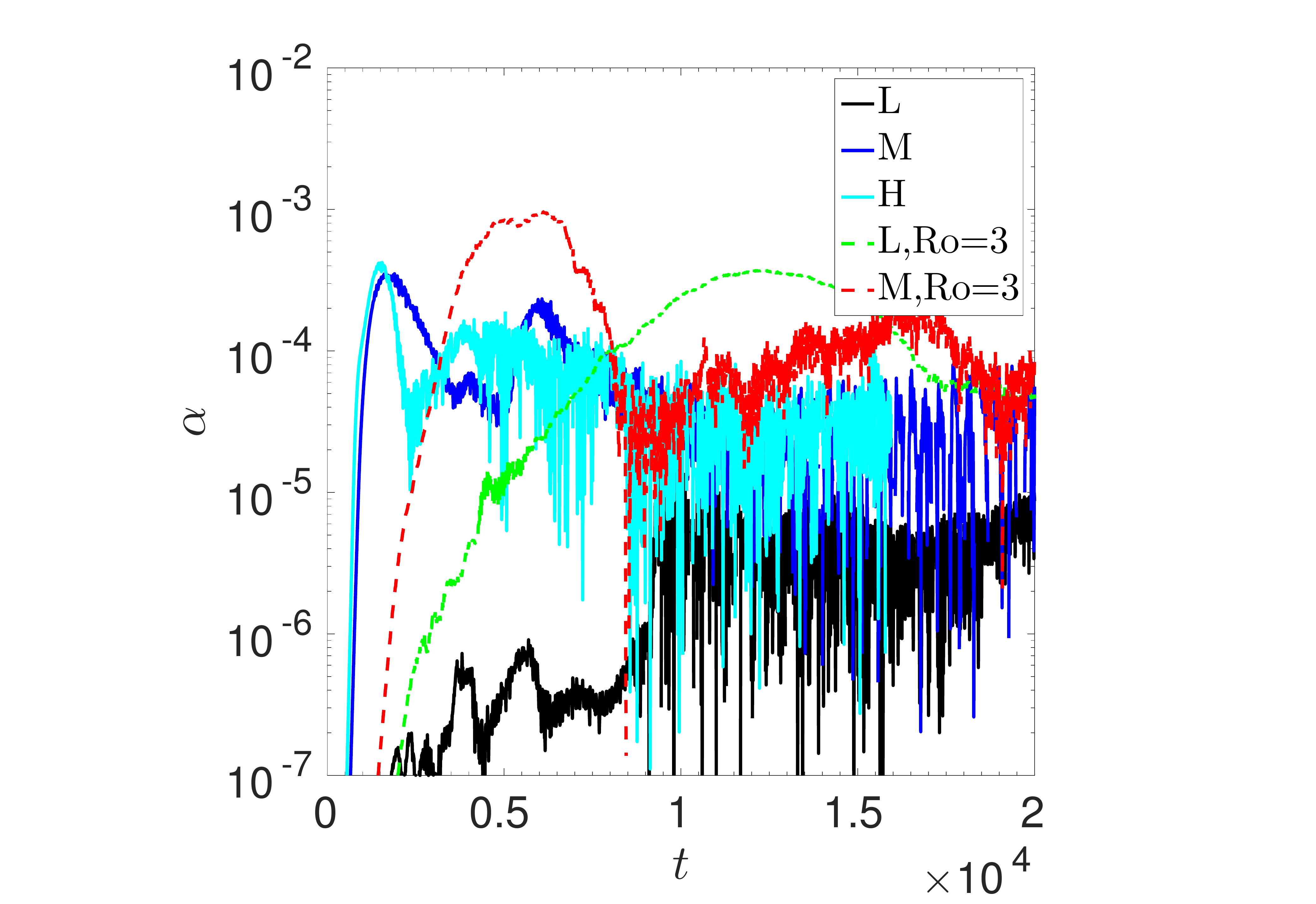}}
    \end{center}
  \caption{Top: RMS turbulent velocities normalised by $c_s$ ($\langle u/c_s\rangle$), which are subsonic with $\sim 0.05 c_s$. Bottom: angular momentum transport $\alpha$ driven by the instability, which exhibits mean values of $\alpha \sim 10^{-5}-10^{-4}$ for our configuration with rigid boundaries.}
  \label{PP2}
\end{figure}

While our primary aim is to study the evolution of eccentric discs, this instability (which is an artifact of our model) will drive subsonic wave activity that will interact with the eccentric mode. Because of this, we briefly describe the properties of this instability for a circular disc in this section, before discussing the evolution of eccentric discs in the next. In Figs.~\ref{PP1} and \ref{PP1a}, we present calculations that show the energy in the lowest ten azimuthal wavenumbers
\begin{eqnarray}
\label{Em}
E_m(t)=\int_{R_i}^{R_o}\frac{1}{2}|\hat{u}_{m}(R,t)|^2 R \, \mathrm{d}\,R,
\end{eqnarray}
where
\begin{eqnarray}
\label{um}
\hat{u}_{m}(R,t)=\frac{1}{\pi}\int_0^{2\pi}u_{R}(R,\phi,t)\mathrm{e}^{-\mathrm{i} m\phi} \mathrm{d}\,\phi,
\end{eqnarray}
which is defined so that it represents the power in the solution with a given value of $|m|$, so that only $m>0$ need to be considered for this quantity. We compare three different numerical resolutions $N_R\times N_\phi$: $200\times 256$ (labelled as L), $400\times 512$ (labelled as M) and $800\times 1024$ (labelled as H). Each simulation has $R_o=2$, $c_s=0.05$, and is started with random white noise perturbations to each component of the velocity on the grid-scale\footnote{No attempt is made to ensure that these are identical for each resolution adopted.} with amplitude $10^{-5}c_s$. Since this instability preferentially excites high azimuthal wavenumbers and there is no explicit viscosity in our simulations, we expect some dependence on resolution, though we will show that the results for M and H appear to have converged. 

Fig.~\ref{PP1} illustrates that simulation L exhibits instability with the excitation of $m\sim 5$ propagating waves by $t\sim 2000$, but this initially saturates at low amplitude. At later times, $m\sim 3$ standing waves develop throughout the domain. We illustrate the velocity field during these stages in Fig.~\ref{PP3} for this simulation, in the early linear growth phase (top) and during the later phase (bottom). The RMS radial velocity driven by the instability, normalised by the sound speed, 
\begin{eqnarray}
\langle u/c_s \rangle = \sqrt{\frac{1}{c^2_s \pi (R^2_o-R^2_i)}\int_{0}^{2\pi}\int_{R_i}^{R_o} u_R^2 R \,\mathrm{d}R \,\mathrm{d}\phi},
\end{eqnarray} 
is plotted in Fig.~\ref{PP2} for all three simulations, along with the efficiency of angular momentum transport 
\begin{eqnarray}
\alpha = \frac{1}{c_s^2\pi (R^2_o-R^2_i)}\int_{0}^{2\pi}\int_{R_i}^{R_o} u_R (u_\phi-R\Omega) R \,\mathrm{d}R \,\mathrm{d}\phi.
\end{eqnarray}
During simulation L , the maximum RMS velocities attained are very subsonic, with $\langle u\rangle\sim 10^{-3}-10^{-2}c_s$, and the angular momentum transport is weak, with $\alpha\lesssim 10^{-5}$. 

Simulations M and H reach larger turbulent velocities of $\langle u\rangle\sim 0.03 c_s$ and somewhat more efficient angular momentum transport with $\alpha\sim 10^{-4}$ (Fig.~\ref{PP2}) -- note that $\alpha=10^{-4}$ implies $t_\mathrm{visc}=\frac{R_i^2}{\alpha c_s H}=\frac{1}{\alpha c_s^2}\approx 4\times 10^{6}$. The initial instability in these simulations preferentially excites $m\sim 9$ components, with $m\sim 6$ dominating at later times, which differs from simulation L. However, the instability appears to be well captured with the resolution adopted for simulation M, since simulation H does not differ significantly. In addition, results do not appear to be strongly dependent on the domain size, with simulations in a domain with $R_o=3$ (where resolutions L and M correspond with $300\times256$ and $600\times 512$ grid-points, respectively) giving similar spectra at late times (compare Fig.~\ref{PP1} with Fig.~\ref{PP1a}) and broadly comparable turbulent velocities and angular momentum transport (Fig.~\ref{PP2}) -- note that the initial turbulent stages are different, with $m\sim 6$ modes excited initially that remain dominant at late times. (We have also checked that the outcome of the instability does not depend significantly on $\sigma$, at least where $\sigma \leq 1.5$.)

The instability excites spiral density waves that saturate with subsonic velocities and lead to weak but non-negligible angular momentum transport. This instability provides background wave activity which complicates the analysis of eccentric modes in the next section. Depending on the particular waves excited by this instability, their nonlinear interaction with the predominantly $m=1$ eccentric mode could either drain energy from, or transfer energy into, this component. Hence, we would not expect secular theory to be valid in the presence of a strong non-axisymmetric component, given that it neglects wave-wave couplings.

Given that this instability is minimised for the resolution adopted for simulation L, we primarily focus on simulations using this resolution in the next section, where we turn to analyse the long-term evolution of eccentric discs.

\subsection{Long-term nonlinear evolution of free eccentric modes}
\label{longterm}

As in \S~\ref{Precession}, we initialise the flow with an eccentric mode with peak eccentricity amplitude $A$ and study its nonlinear evolution, focussing on its behaviour over many ($50-70$) precession periods.

\begin{figure}
  \begin{center}
     \subfigure{\includegraphics[trim=6cm 0cm 8cm 0cm, clip=true,width=0.45\textwidth]{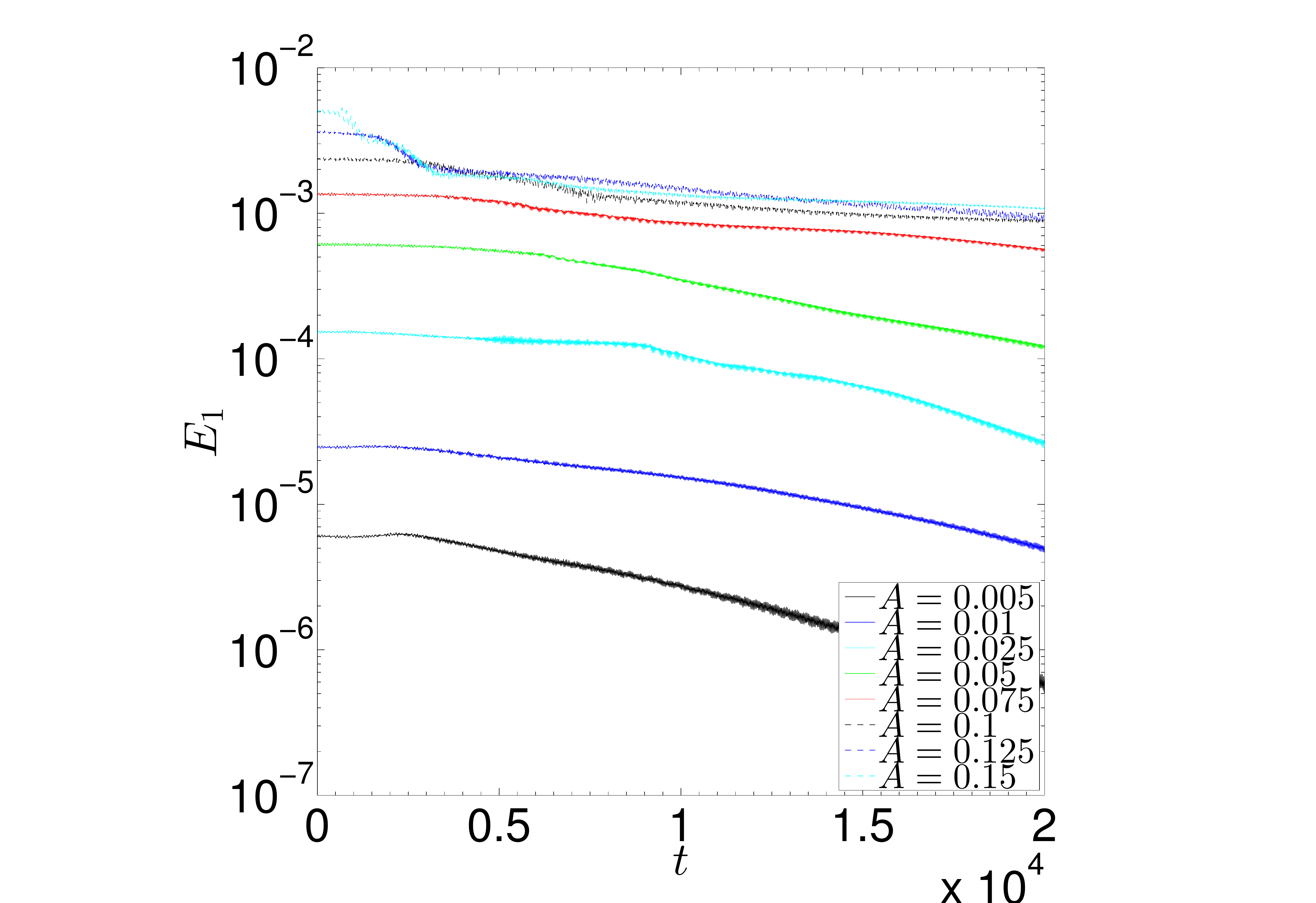} } 
    \end{center}
  \caption{Energy in the eccentric mode $E_1$ (technically the energy in the $|m|=1$ components of the flow) as a function of time for a set of simulations with $R_o=2$, $c_s=0.05$, $\sigma=0$ for varying $A$ (each with a resolution of $200\times 256$). The simulation is run for approximately 70 linear precession periods and shows that the eccentric mode persists throughout the duration of these simulations. It does, however, experience gradual amplitude-dependent damping.}
  \label{0}
\end{figure}

\begin{figure}
  \begin{center}
     \subfigure[$A=0$]{\includegraphics[trim=6.5cm 0cm 7cm 0cm, clip=true,width=0.23\textwidth]{2a}}
     \subfigure[$A=0.005$]{\includegraphics[trim=6.5cm 0cm 7cm 0cm, clip=true,width=0.23\textwidth]{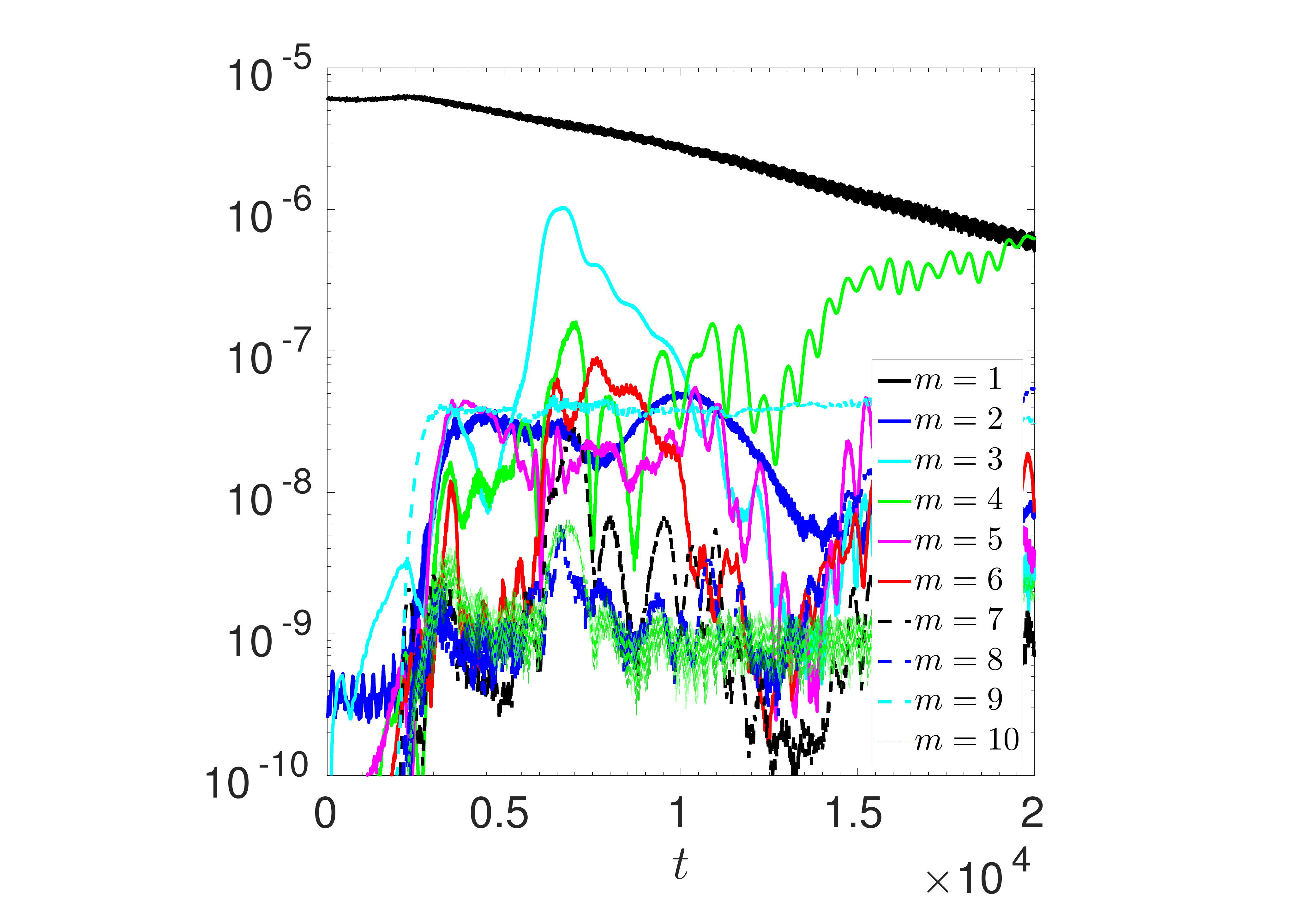} } 
     \subfigure[$A=0.01$]{\includegraphics[trim=6.5cm 0cm 7cm 0cm, clip=true,width=0.23\textwidth]{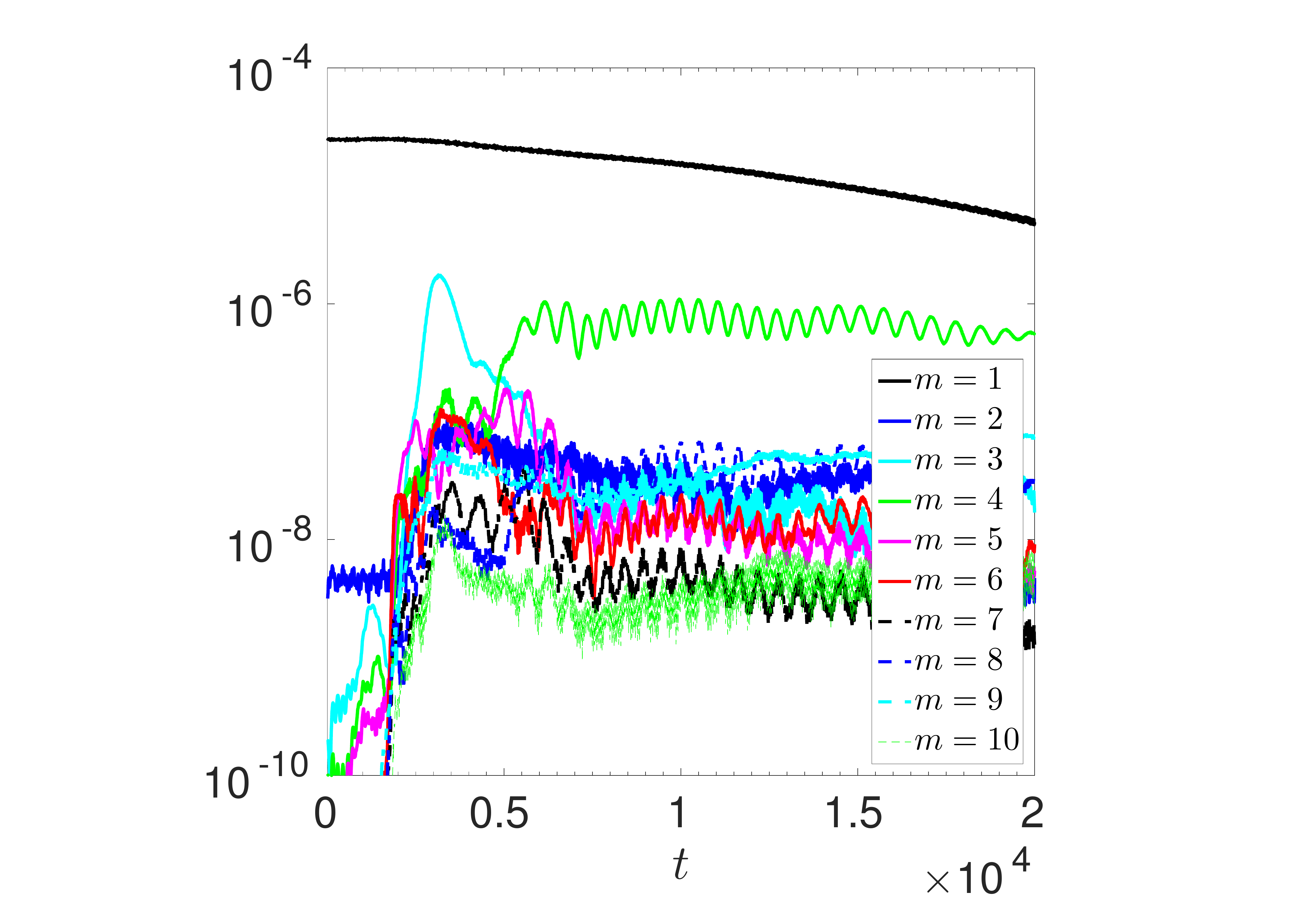} } 
     \subfigure[$A=0.05$]{\includegraphics[trim=6.5cm 0cm 7cm 0cm, clip=true,width=0.23\textwidth]{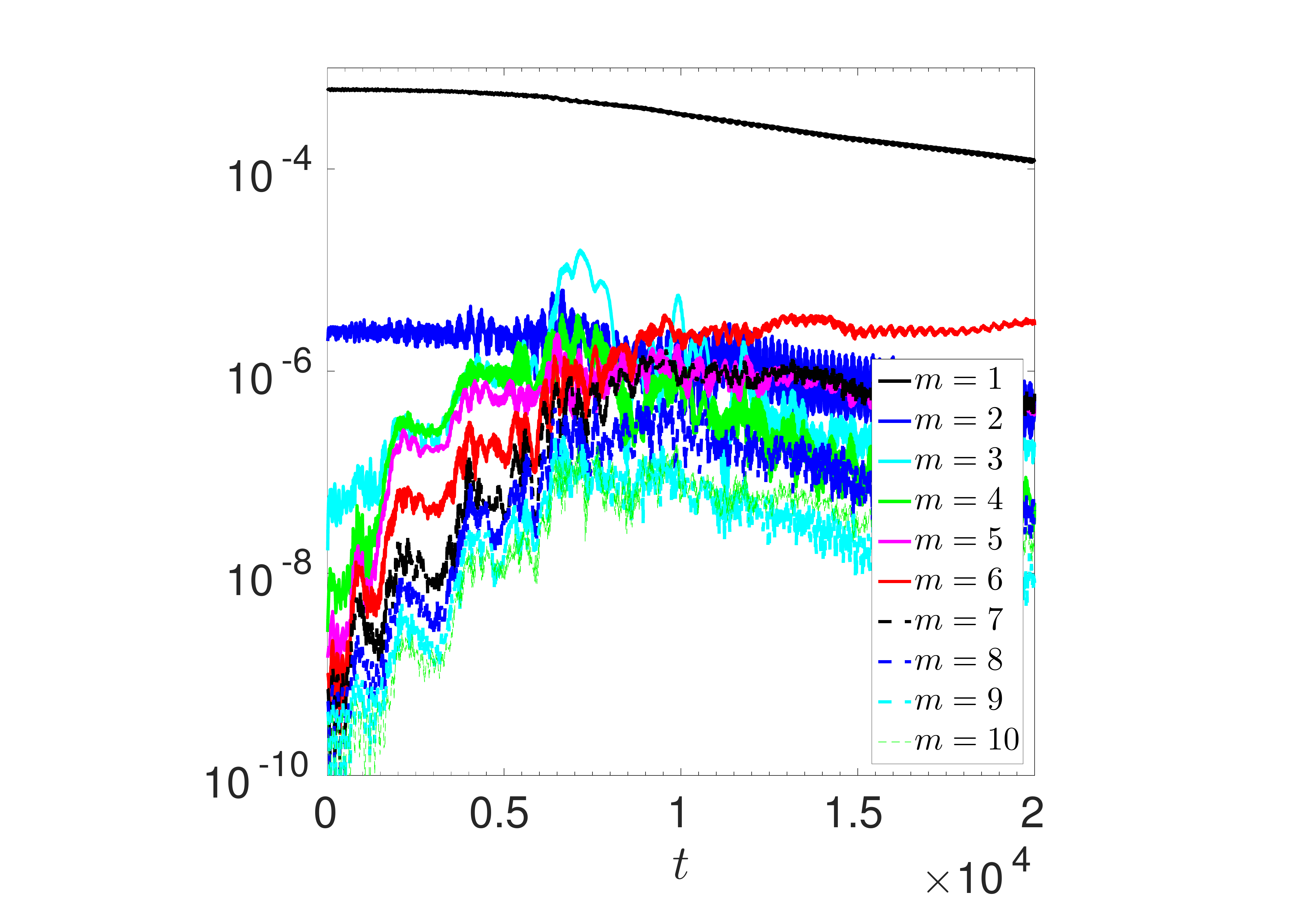} } 
     \subfigure[$A=0.1$]{\includegraphics[trim=6.5cm 0cm 7cm 0cm, clip=true,width=0.23\textwidth]{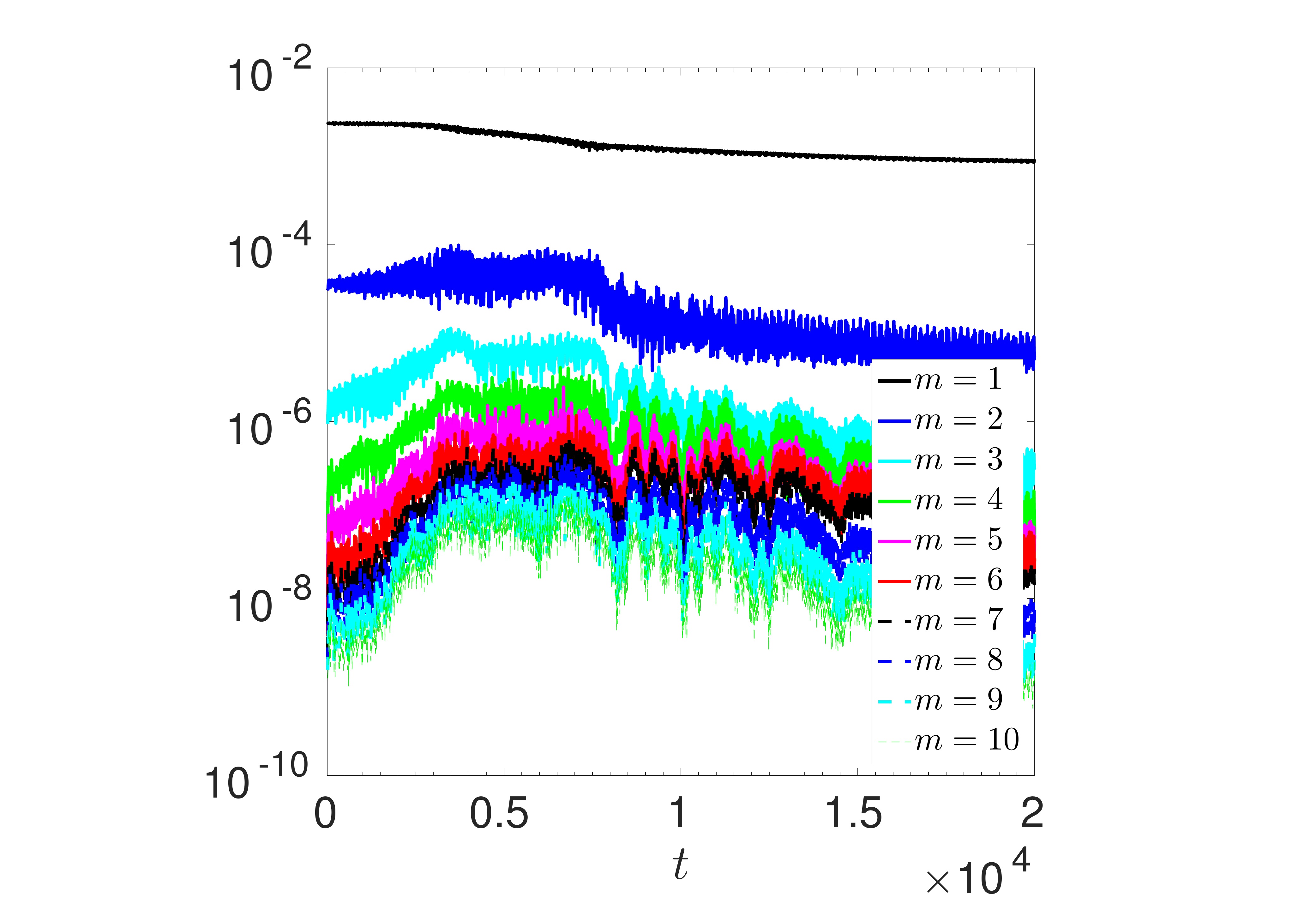} } 
     \subfigure[$A=0.15$]{\includegraphics[trim=6.5cm 0cm 7cm 0cm, clip=true,width=0.23\textwidth]{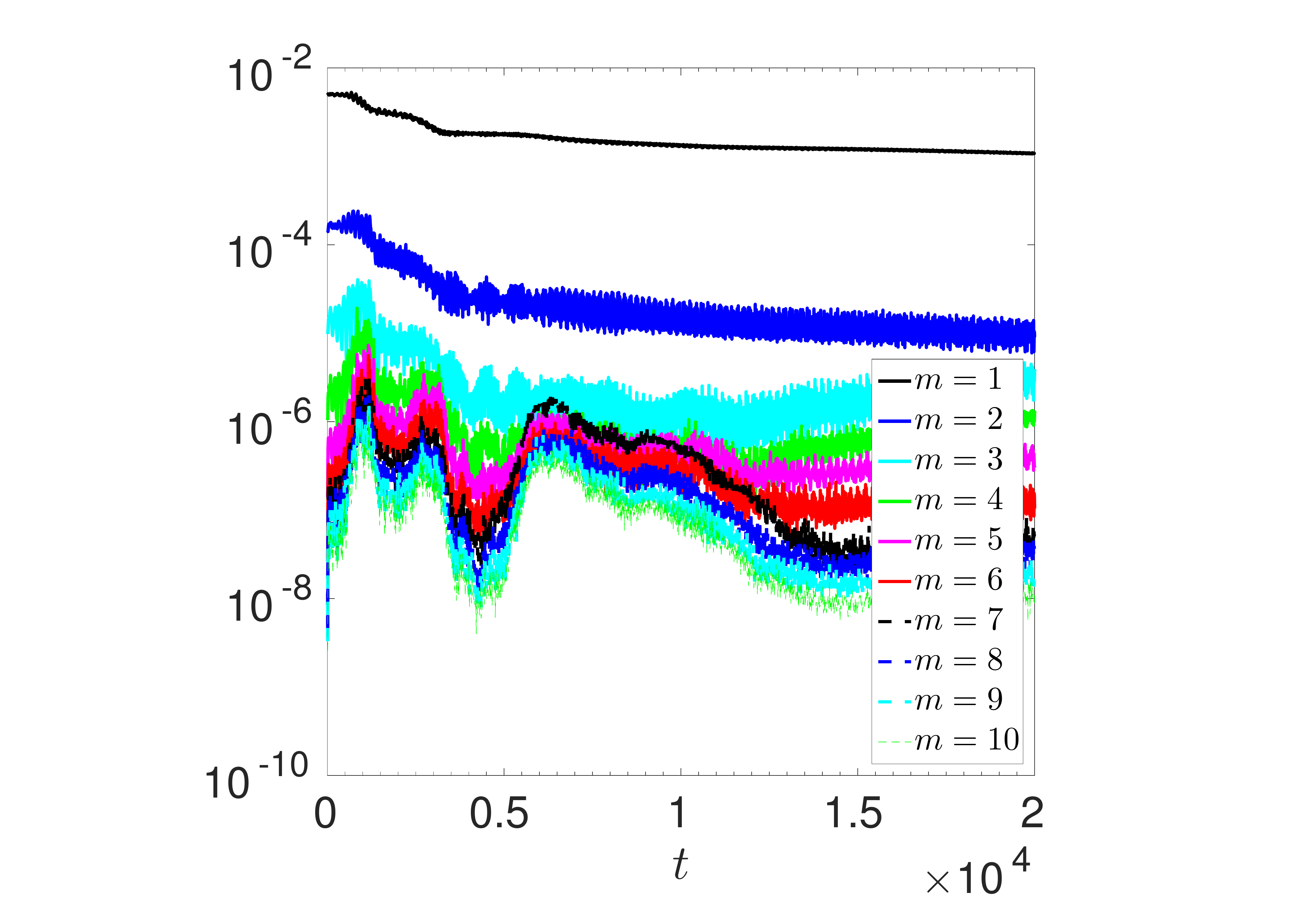} } 
   \end{center}
  \caption{Energy in different azimuthal wavenumber components of the solution ($E_m$) for $m\in [1,10]$ as a function of time for several different amplitudes $A$ for a set of simulations with $200\times 256$ grid points. The instability that occurs in the absence of an eccentric mode also occurs when $A\ne 0$, which excites non-axisymmetric waves that subsequently interact with the eccentric mode, generally leading to damping. However, the eccentric mode itself does not appear to be subject to separate instabilities in two dimensions.}
  \label{1}
\end{figure}

We begin by analysing the temporal evolution of the eccentric mode energy for various $A$ in Fig.~\ref{0}. We plot $E_1$ (as defined in Eq.~\ref{Em}), which approximately represents the energy in the eccentric mode even for moderate $A$, when $m\ne1$ components are also present (see Fig.~\ref{1}). The validity of secular theory is indicated by the lack of evolution in $E_1$ during the initial stages. The eccentric mode persists throughout our simulations, but experiences two different kinds of damping as the simulation progresses. Firstly, numerical diffusion acts to damp the eccentric mode, which is particularly pronounced for cases with higher $A$. Since the streamlines are then highly concentrated (see e.g.~the bottom panel of Fig.~\ref{secularfig}), and there are strong eccentricity gradients, these modes experience appreciable damping by numerical diffusion on the grid-scale. However, this is only the most important damping mechanism during the initial stages when $A\gtrsim 0.125$. During later stages (i.e.~after $t\gtrsim 1000$), the dominant damping mechanism is the interaction of the $m=1$ eccentric mode with other non-axisymmetric waves that are driven by the instability discussed in \S~\ref{PPinst}. These waves are excited in the absence of an eccentric mode, but interact with the eccentric mode through nonlinear interactions to generate additional non-axisymmetric components. In most cases this interaction acts to damp the eccentric mode, but it can also transfer energy into the eccentric mode in some cases e.g.~see $A=0.005$ when $t\sim 2500$. 

\begin{figure}
  \begin{center}
     \subfigure[$A=0$]{\includegraphics[trim=6.5cm 0cm 7cm 0cm, clip=true,width=0.23\textwidth]{2b}}
     \subfigure[$A=0.01$]{\includegraphics[trim=6.5cm 0cm 7cm 0cm, clip=true,width=0.23\textwidth]{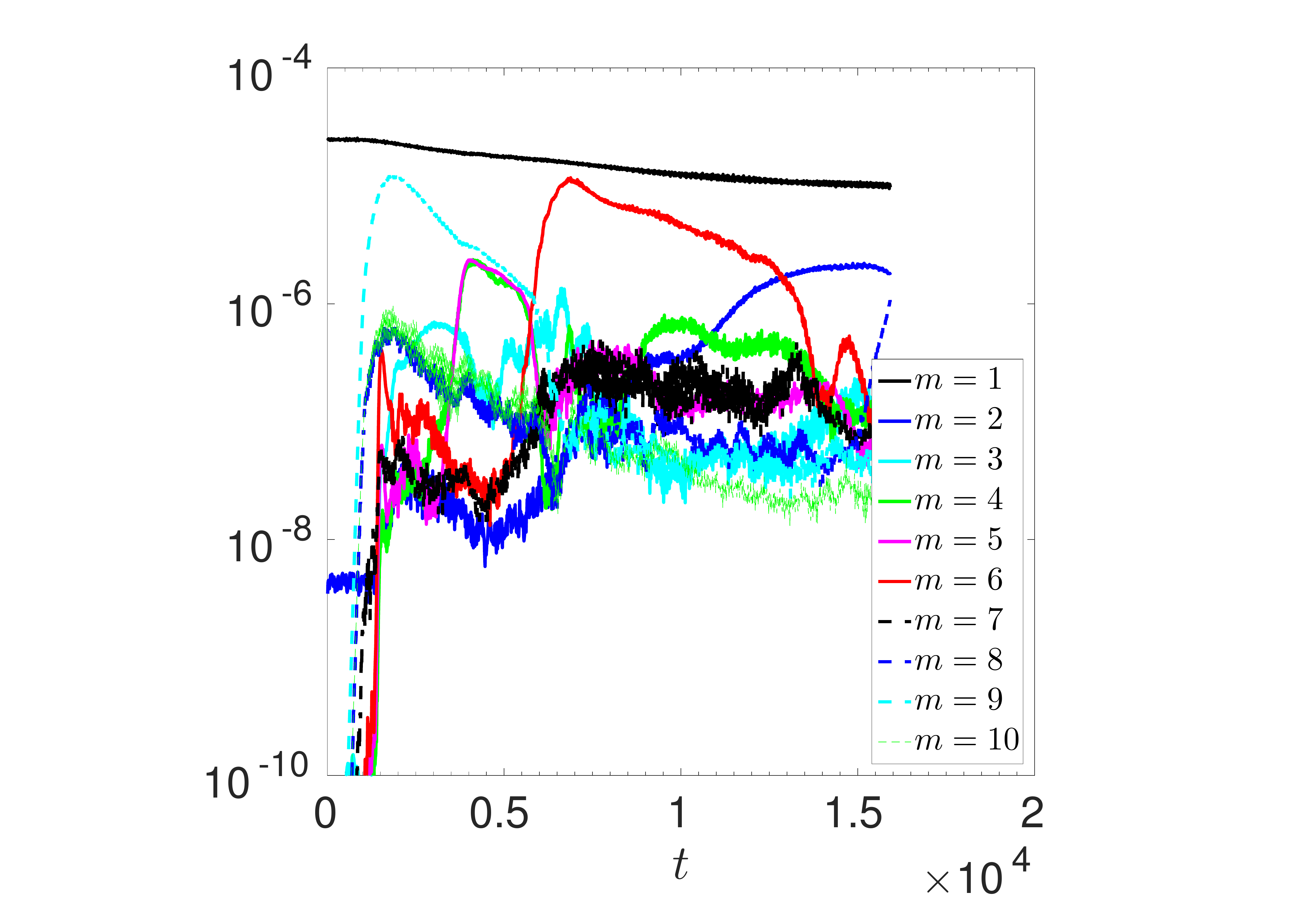} } 
     \subfigure[$A=0.05$]{\includegraphics[trim=6.5cm 0cm 7cm 0cm, clip=true,width=0.23\textwidth]{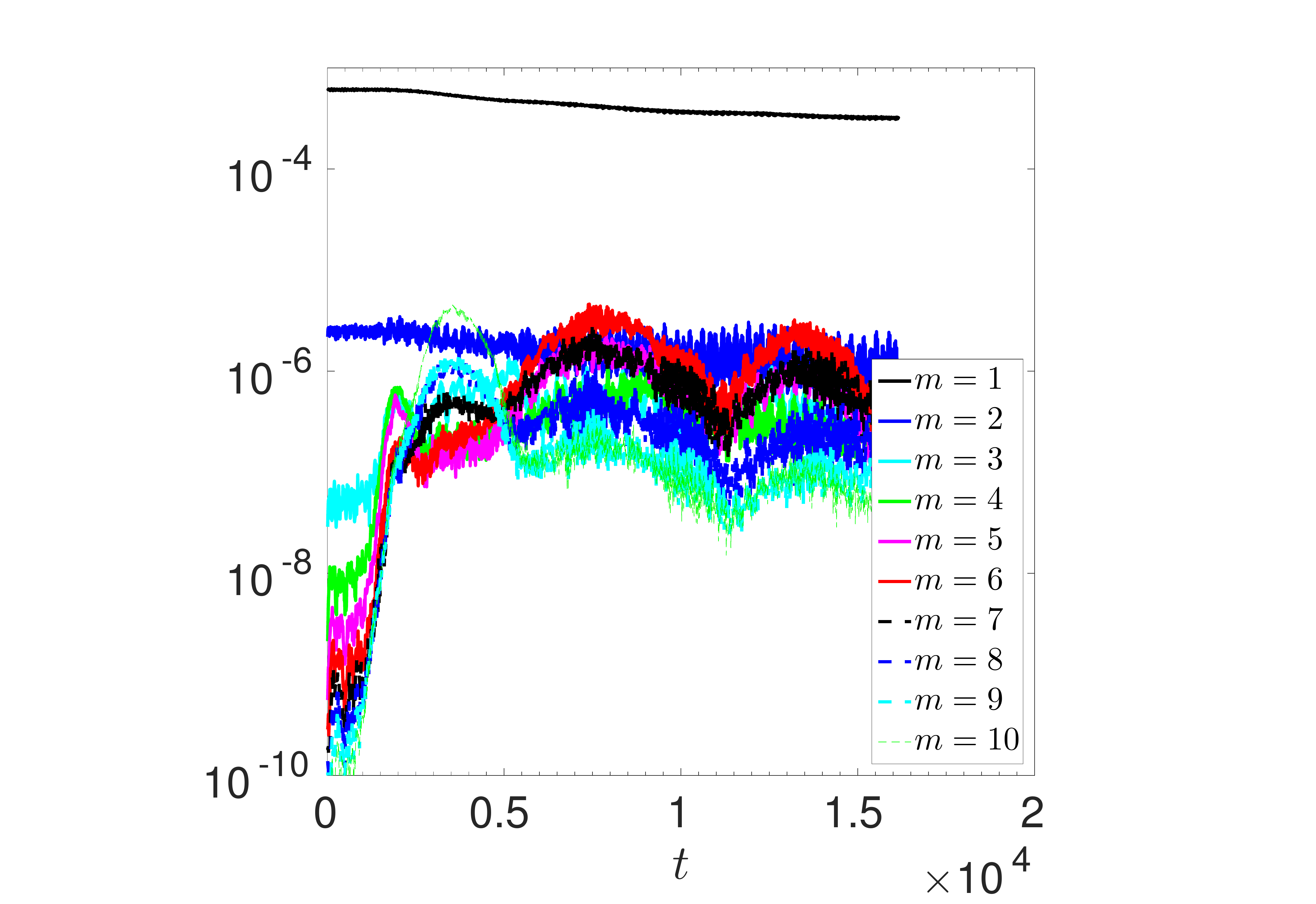} } 
     \subfigure[$A=0.1$]{\includegraphics[trim=6.5cm 0cm 7cm 0cm, clip=true,width=0.23\textwidth]{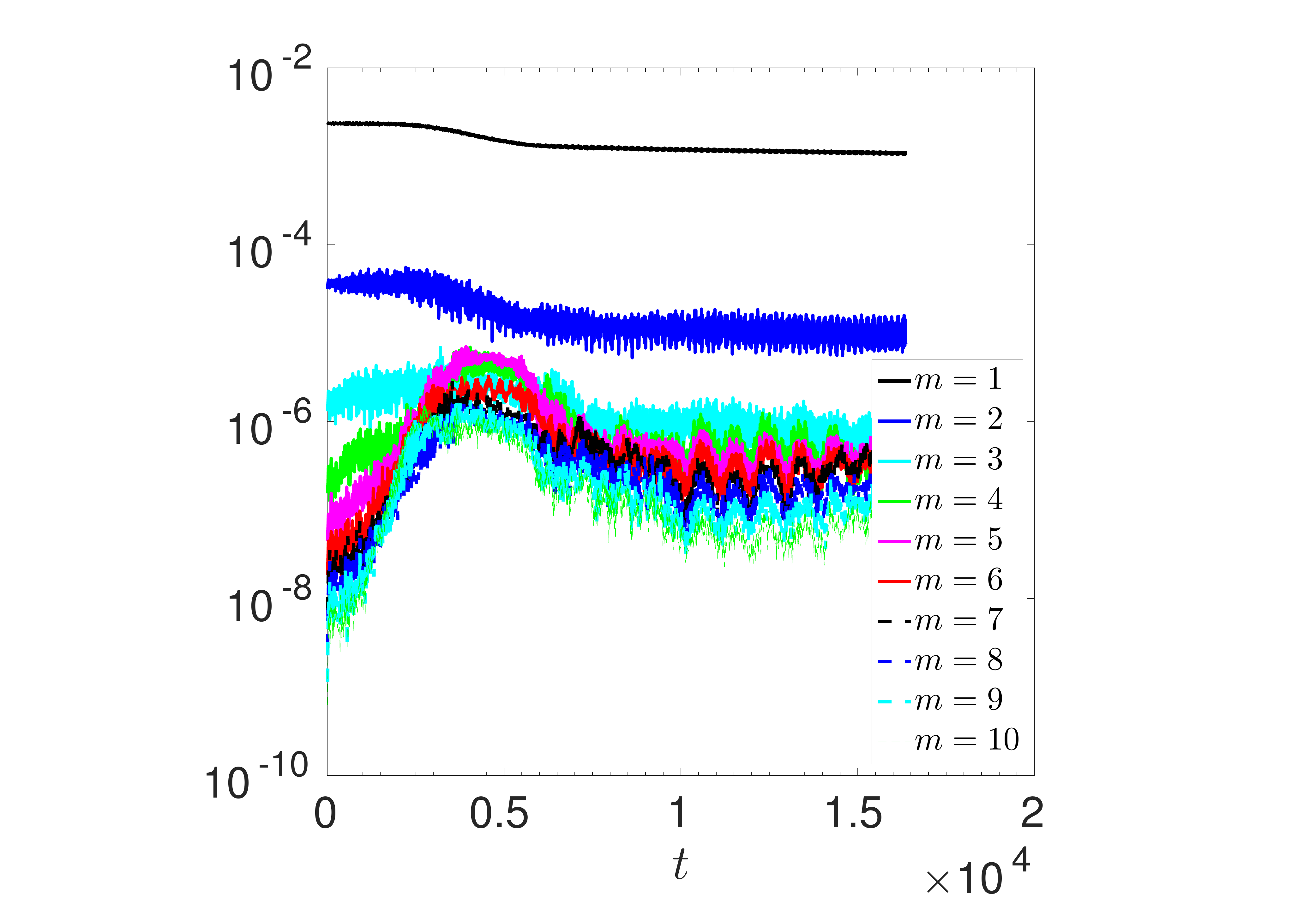} } 
    \end{center}
  \caption{Same as Fig.~\ref{1} except for a set of simulations with $400\times 512$ grid points.}
  \label{1a}
\end{figure}

In Fig.~\ref{1}, we have plotted the temporal evolution of $E_m$ for $m\in[1,10]$ to illustrate the growth of $m\ne 1$ components of the solution. The top left panel repeats our results when $A=0$ (Fig.~\ref{PP1}) to provide a point of reference. The remaining panels show the same quantity for several different values of $A$. When $A=0.005$, background instability occurs at approximately the same time as the $A=0$ case, preferentially exciting similar modes (together with those with azimuthal wavenumbers that differ by 1), and the $m\ne 1$ components of the solution saturate with similar energies ($\sim 10^{-7}$). The $m=1$ component does not undergo significant damping until $t\sim 2500$, after the background instability has set in. This supports our interpretation that this damping of the eccentric mode is due to its interaction with non-axisymmetric waves driven by the background instability, and not due to a separate instability of the eccentric mode itself.

In Fig.~\ref{1a}, we have plotted the same quantity but for a set of simulations that have double the radial and azimuthal resolution (starting with simulation M of \S~\ref{PPinst}). The eccentric mode evolves similarly in these cases as in the lower resolution simulations in Fig.~\ref{1}. However, the background instability is excited earlier in the simulations with small $A$, which leads to slightly different long-term quantitative evolution for the eccentric mode amplitude (in fact there is slightly more efficient damping in some of these simulations compared with those with the lower resolution, demonstrating that the amplitude decay is not primarily due to numerical diffusion).

As $A$ is increased, this instability occurs sooner in the simulation, presumably because there is more energy in $m\ne 1$ components of the solution at $t=0$. However, with the exception of the $m=2$ component (which arises primarily to describe the eccentric mode itself when $A\gtrsim  0.05$), other non-axisymmetric components typically saturate with similar power to the $A=0$ case. In all simulations, there is an amplitude- and time-dependent damping (or transient growth, as briefly observed at $t\approx 2000$ in the $A=0.005$ simulation in Fig.~\ref{1}) of the eccentric mode due to its interaction with waves driven by the background instability. However, our interpretation, which is guided by Figs.~\ref{1} and \ref{1a}, is that the eccentric mode does not appear to be subject to additional instabilities due to the presence of a nonzero eccentricity (at least in two dimensions), only to the finite-$A$ modification of the instability that occurs when $A=0$ due to our adoption of rigid walls.

\begin{figure}
  \begin{center}
     \subfigure[$A=0.01$]{\includegraphics[trim=5cm 0cm 7cm 0cm, clip=true,width=0.23\textwidth]{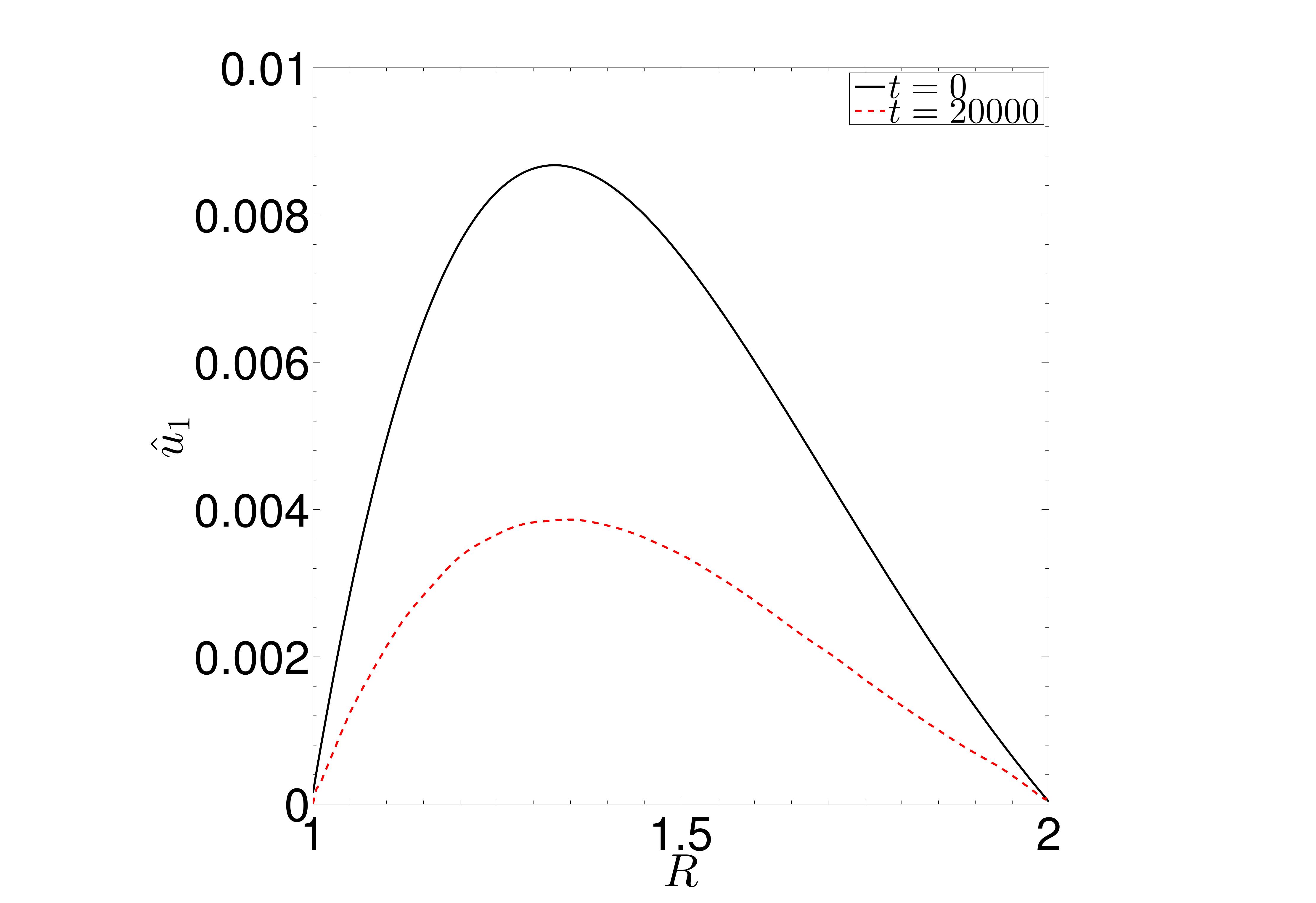} } 
     \subfigure[$A=0.05$]{\includegraphics[trim=5cm 0cm 7cm 0cm, clip=true,width=0.23\textwidth]{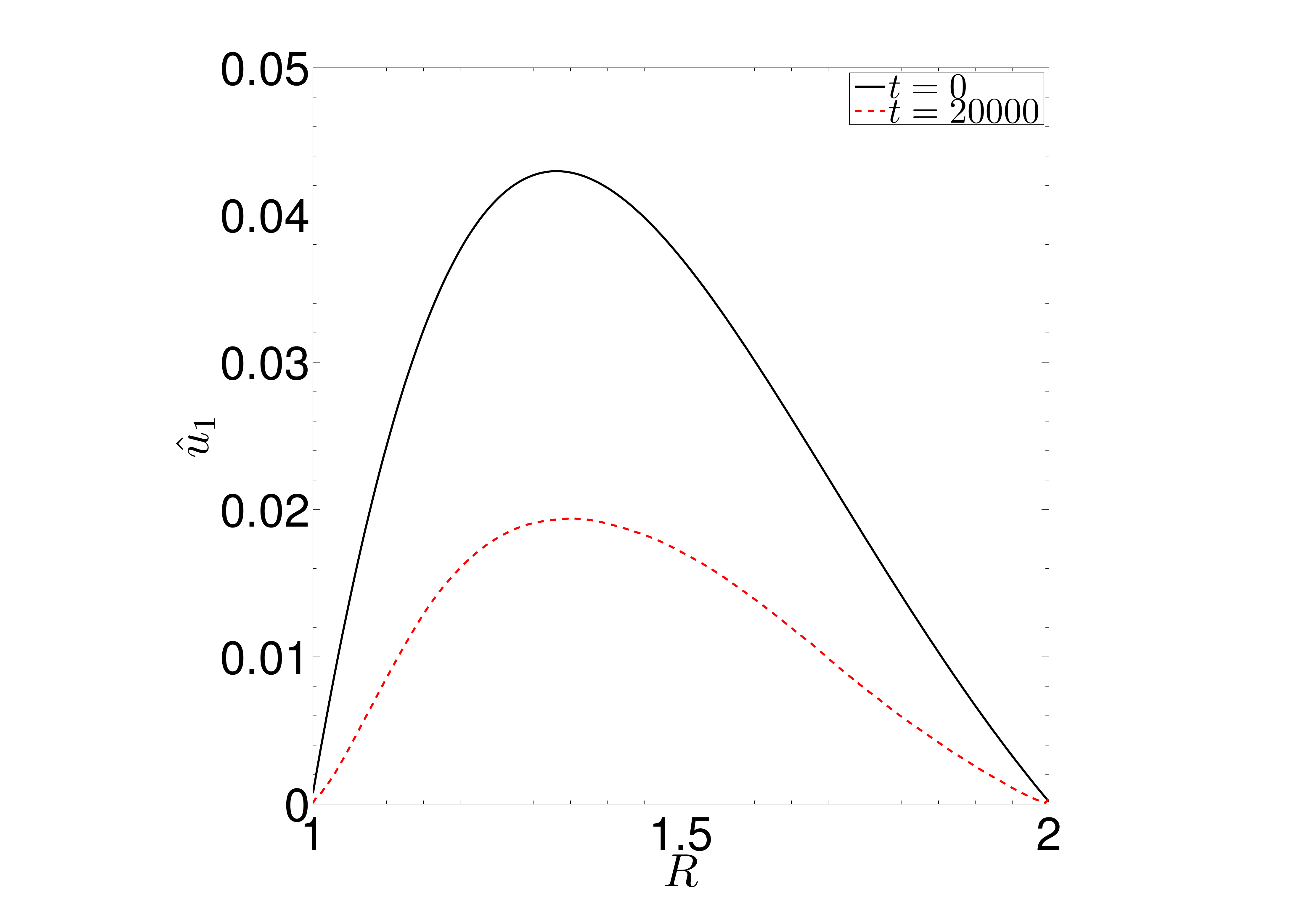} } 
     \subfigure[$A=0.1$]{\includegraphics[trim=5cm 0cm 7cm 0cm, clip=true,width=0.23\textwidth]{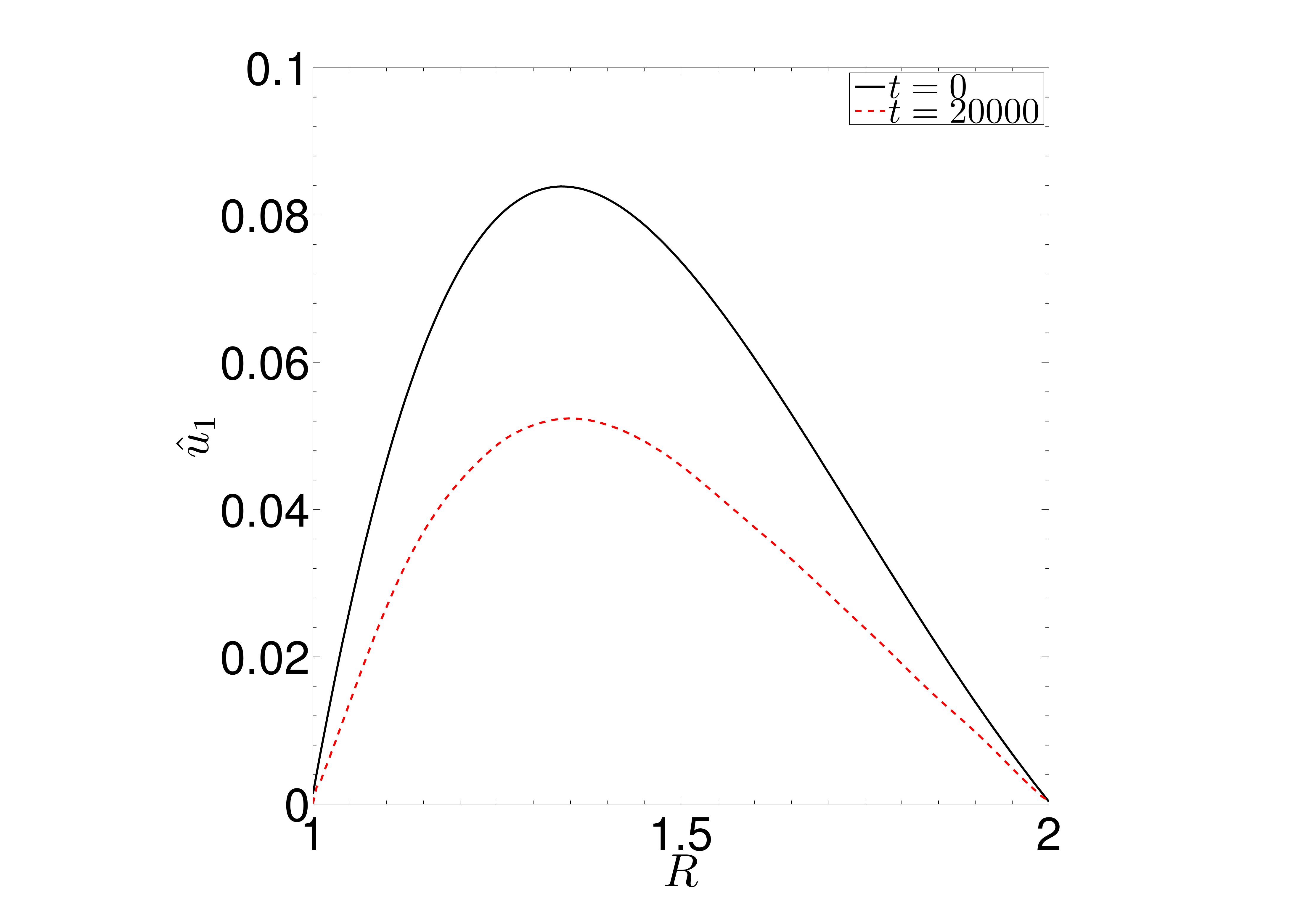} } 
     \subfigure[$A=0.15$]{\includegraphics[trim=5cm 0cm 7cm 0cm, clip=true,width=0.23\textwidth]{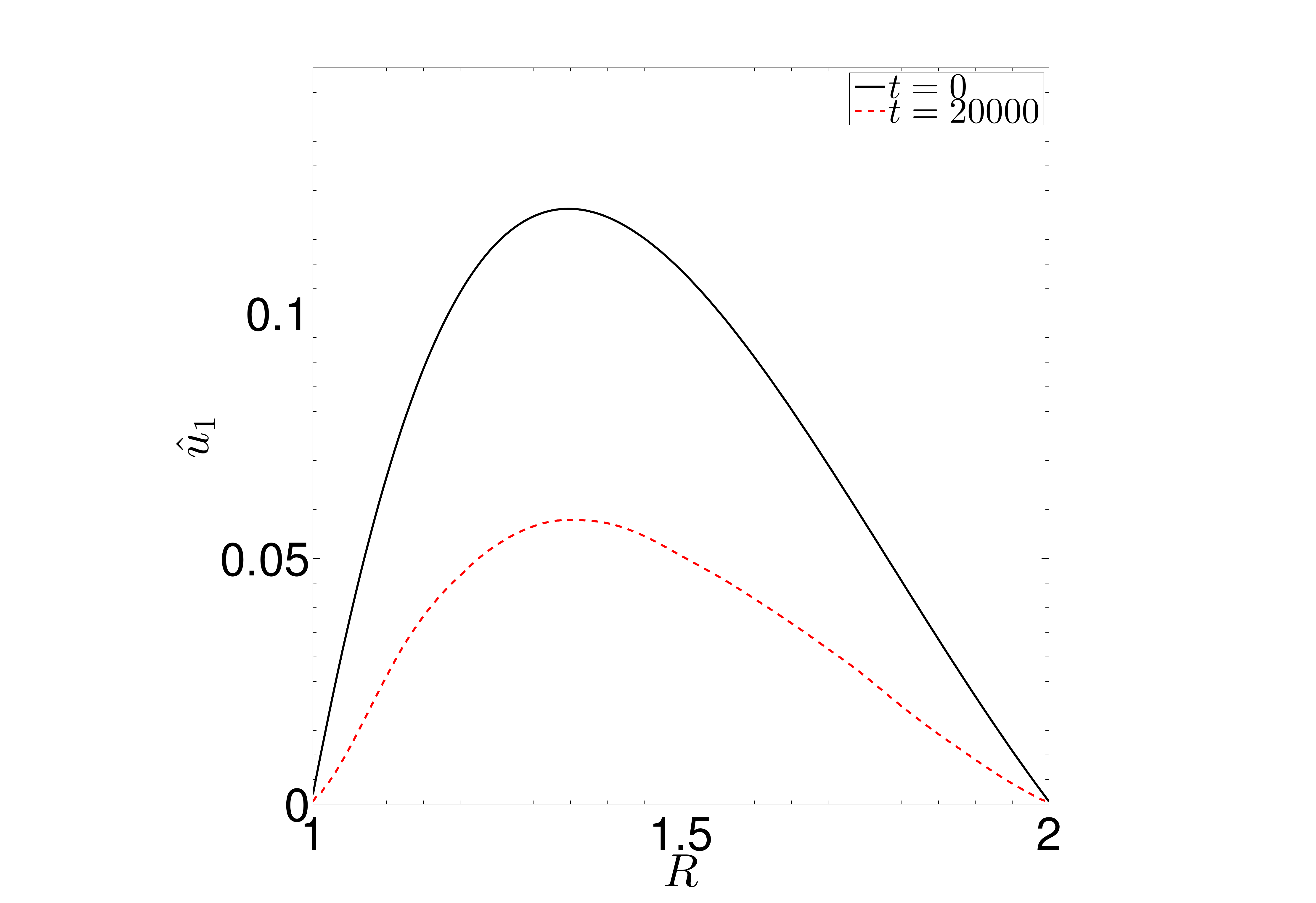} } 
    \end{center}
  \caption{Magnitude of the radial velocity in the $m=1$ component of the solution ($|\hat{u}_1|$) as a function of $R$ at the beginning of the simulation (black solid lines; taken from nonlinear secular theory) and at $t=20000$ (red dashed lines; representing the end of our simulations). This illustrates the damping of the eccentric mode, and also that its shape remains similar to that predicted by secular theory.}
  \label{2}
\end{figure}

\begin{figure}
  \begin{center} 
     \subfigure[$A=0.01$]{\includegraphics[trim=7cm 0cm 7cm 0cm, clip=true,width=0.23\textwidth]{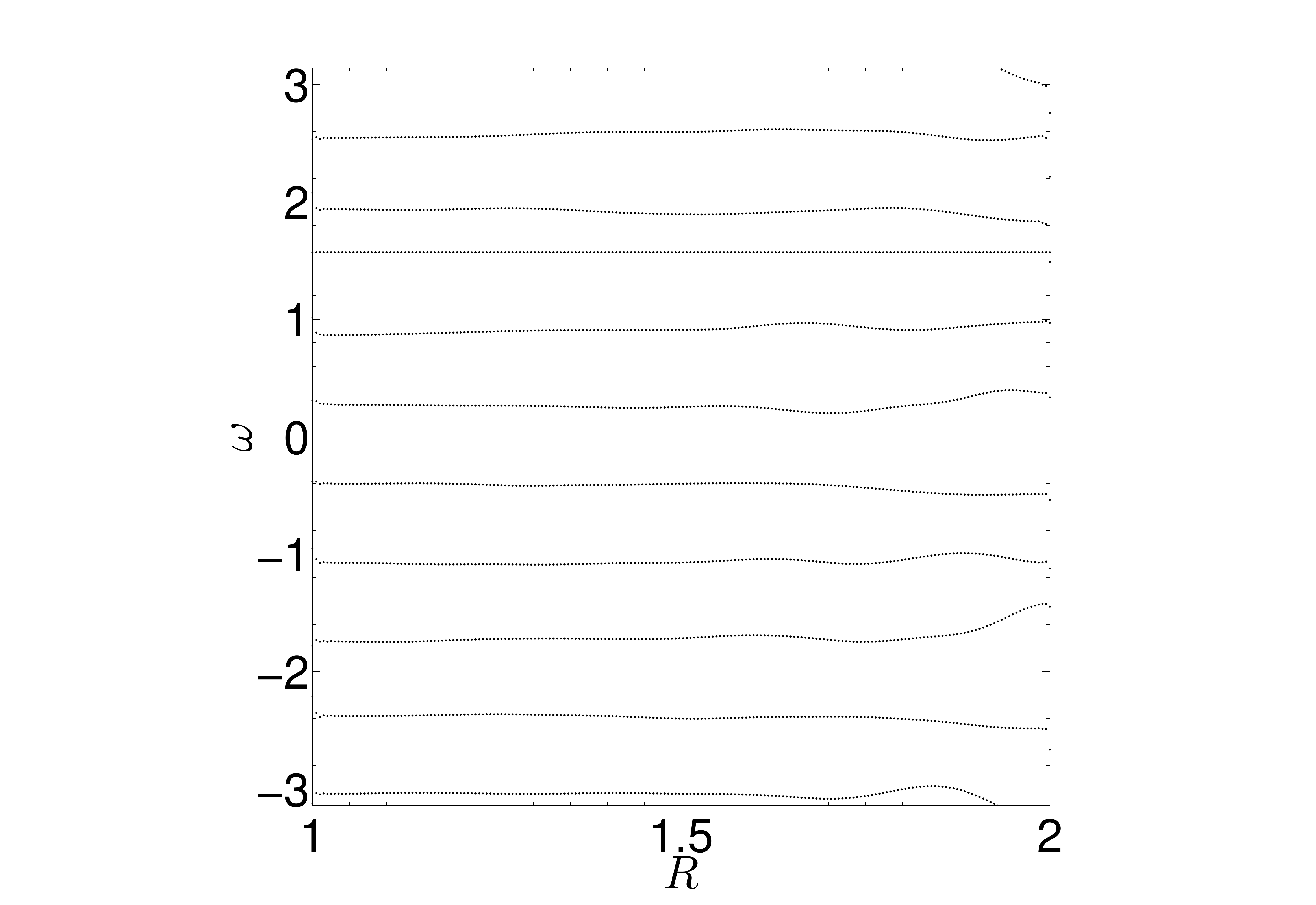} } 
     \subfigure[$A=0.1$]{\includegraphics[trim=7cm 0cm 7cm 0cm, clip=true,width=0.23\textwidth]{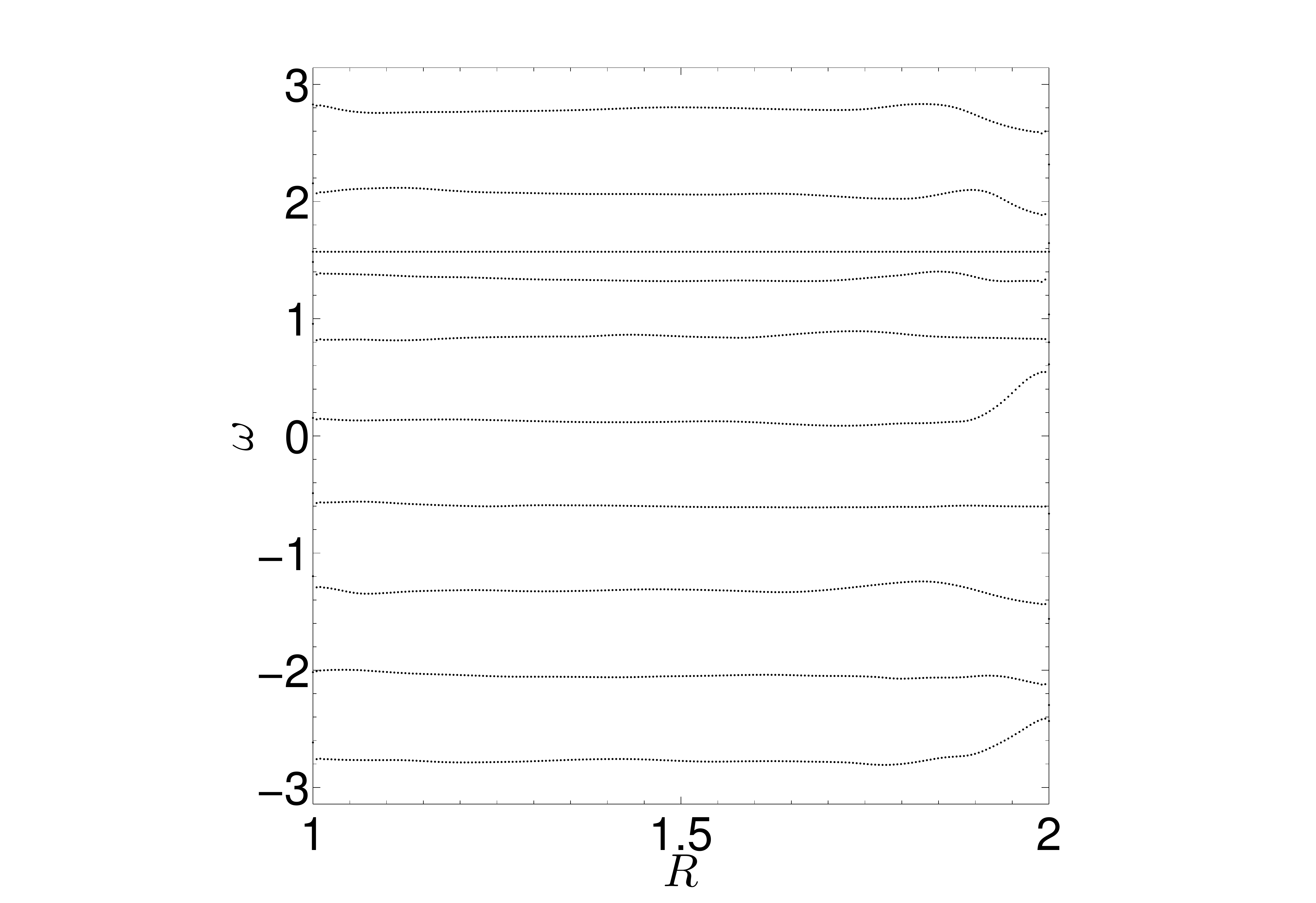} } 
    \end{center}
  \caption{Longitude of pericentre $\omega$ as a function of $R$ for simulations at times $t=[0,30,60,90,120,150,180,210,230,260,290]$. This illustrates that the eccentric disc remains approximately untwisted during one precession period, though slight twists can develop.}
  \label{3}
\end{figure}

In Fig.~\ref{2}, we plot $|\hat{u}_1(R)|$ at $t=0$ (solid black lines) and $t=20000$ (red dashed lines; marking the end of each simulation), which shows the damping of the eccentric mode. In addition, this demonstrates that the eccentric mode remains coherent, persisting throughout the duration of these simulations in a form that is similar to the initial conditions. The similarity in the mode shape at these two times supports the validity of secular theory in describing the shapes of these modes.

Our initial disc model is untwisted, but we do not constrain it to remain so. In Fig.~\ref{3}, we show $\omega(R)$ for $A=0.01$ and $A=0.1$, for times $t\in[0,300]$ (slightly more than $P_p$) spaced in intervals of 30 time units. At $t=0$, the disc is untwisted with $\omega=\pi$, but slight twists develop during the course of the simulation, preferentially near to the outer boundary. However, the disc remains approximately untwisted as it evolves throughout these simulations with a net twist that is small even when $A=0.1$ (smaller than 0.5 rads). It is possible that numerical damping, which is likely to act in a similar way to a shear viscosity \citep{Ogilvie2001}, could be responsible for this twist in the outer regions (where the grid cells are largest), thereby preventing the disc from remaining entirely untwisted (also, if the initial conditions are not exact nonlinear solutions, some twist would be expected to develop). Alternatively, since $e\rightarrow 0$ as $R\rightarrow 2$, the phase of the complex eccentricity becomes undefined at this location, so we might expect the twist at this location to be arbitrary.

\begin{figure}
  \begin{center}
     \subfigure{\includegraphics[trim=5cm 0cm 5cm 0cm, clip=true,width=0.4\textwidth]{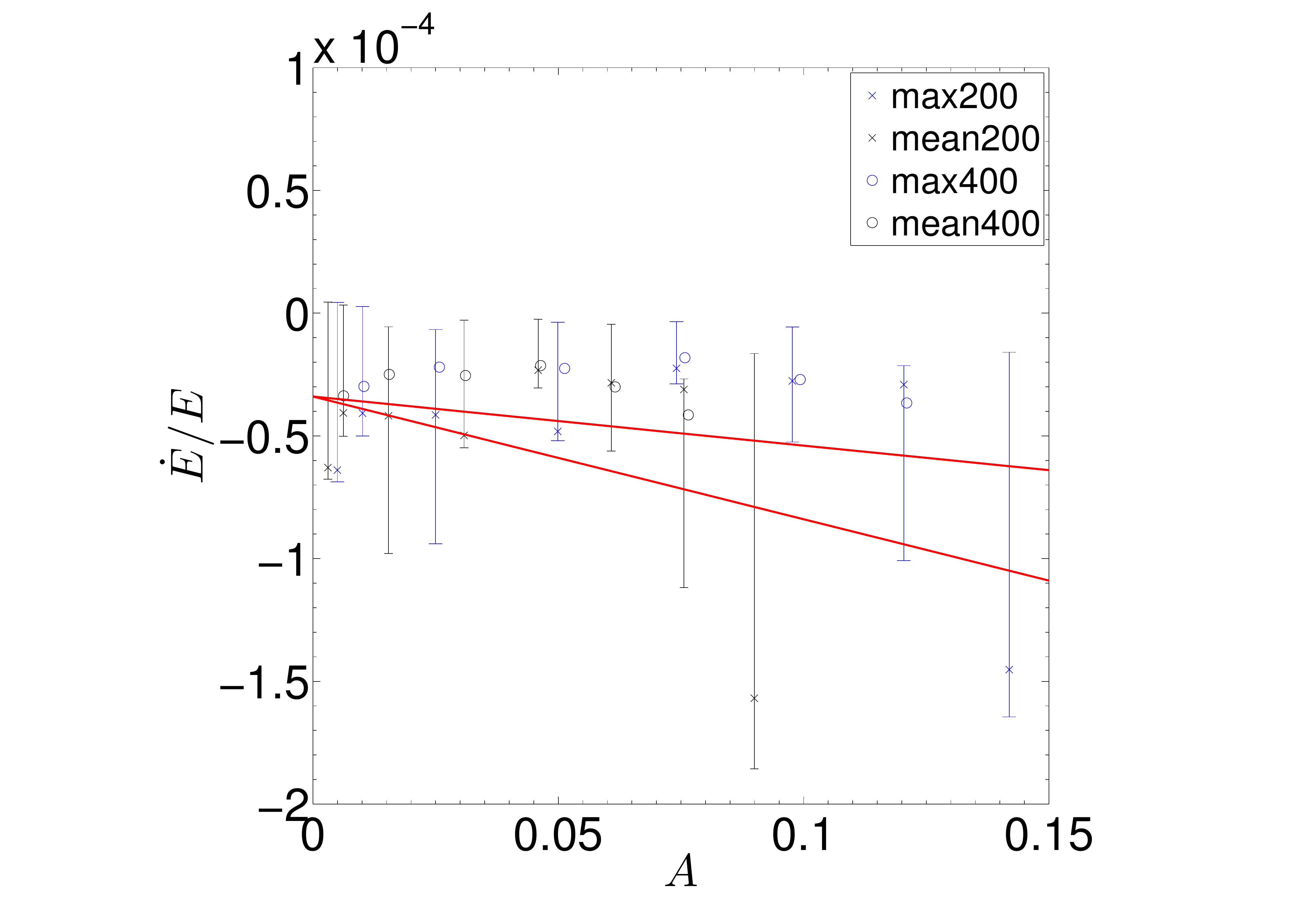} }
    \end{center}
  \caption{Decay rate of maximum and mean eccentricity in the eccentric mode $\dot{E}/E$, as a function of $A$ at resolutions of $200\times 256$ and $400\times 512$. The error bars represent the maximum and minimum damping rates observed in these simulations. The red lines are plotted for reference, and have approximate slopes of $-2\times 10^{-4}$ and $-5\times 10^{-4}$, respectively. Note that $\dot{E}/E\approx -5\times 10^{-5}$ corresponds with a damping time of approximately $2\times 10^4$.}
    \label{4}
\end{figure}

Finally, we analyse the decay rate of the eccentric mode, which we plot in Fig.~\ref{4} for several simulations, showing the effects of varying the resolution. This picture is not clear-cut because the background instability is stronger for the higher resolution case. This illustrates that eccentric modes in two dimensions persist for a very long time, even in the presence of amplitude-dependent wave-wave interactions and numerical damping. 

\subsection{Summary}

In this section we have presented the results of nonlinear hydrodynamical simulations designed to study the long-term evolution of eccentric discs in two dimensions. Eccentric discs remain coherent and do not appear to be subject to instabilities as a result of their free eccentricity in two dimensions, and would presumably persist forever if there was no background instability. They are, however, damped (typically) by their nonlinear interaction with non-axisymmetric spiral density waves -- these waves are driven by a background (``Papaloizou-Pringle") instability in our setup with rigid walls\footnote{This instability also occurs (with a similar growth rate) with free boundaries, and we would expect its nonlinear evolution to be only weakly dependent on the boundary conditions, unless wave reflection from the boundaries is inhibited.}. In real discs there are various mechanisms that could produce such an incoherent ensemble of non-axisymmetric waves e.g.~additional sources of turbulence such as magneto-rotational instability (e.g.~\citealt{HP2009a,HP2009b}), gravitational instability (e.g.~\citealt{PapSav1991,Laughlin1996}), convection (e.g.~\citealt{MR2011}), or possibly the tidal interaction between multiple proto-planets and the disc.

Previous simulations of eccentric modes by \cite{John2005b} used linear secular theory to construct the initial conditions. This enhances the damping of the eccentric mode due to the generation of shocks at the inner boundary of the domain, since we have observed the linear mode to cause orbital intersections if $A$ is sufficiently large. On the other hand, the modes that we have input using nonlinear secular theory do not experience such strong shock-induced damping, as indicated in Fig.~\ref{0}. These modes do experience amplitude-dependent damping by numerical diffusion, which can be particularly strong for large $A$ cases, where the orbits are closer to intersecting. But in our simulations the most important damping mechanism is the interaction of the eccentric mode with other non-axisymmetric waves. 

\section{Conclusions}

In this paper we have studied fundamental aspects of eccentric gas discs that should have general applicability.
We have derived analytically a nonlinear secular theory for isothermal two-dimensional untwisted eccentric Keplerian discs (valid for arbitrary eccentricities and eccentricity gradients for which neighbouring orbits do not intersect; Appendix~\ref{nonlineartheory}), 
and verified its predictions with idealised hydrodynamical simulations using the PLUTO code. Linear secular theory is found to accurately describe the structures and precession rates of eccentric discs with small eccentricities (and eccentricity gradients), which precess in a retrograde sense due to gas pressure. Discs with larger eccentricities (and eccentricity gradients) are observed to precess at a faster rate, which we have explained as a modification of the pressure forces, and resulting disc structure to prevent orbital intersections, that occurs when the orbits in a disc nearly intersect.

The nonlinear modification of the pressure forces might be particularly important for the highly eccentric discs produced in tidal disruption events \citep{Guillochon2011,Liu2013,Guillochon2014,CoughlinNixon2015}, or for narrow eccentric gas rings (cf.~\citealt{BGT1983}, which was applied to planetary rings). Another potential application is to the period excess in superhump binary systems, which is thought to be explained due to the precession of an eccentric gaseous disc \citep{Whitehurst1988,Lubow1991a,Murray2000,Goodchild2006,Smith2007}. However, there is observational evidence of temporal variability of the period excess \citep{Mason2008,Nakata2014}, which may in part be caused by evolution of the pressure forces due to the varying eccentricity and eccentricity gradients in the disc. It would be worthwhile to apply nonlinear secular theory to this problem (e.g.~extending \citealt{Goodchild2006}) in order to explore this possibility further.

Eccentric discs do not appear to exhibit hydrodynamic instability as a result of their free eccentricity in two dimensions, and we would expect them to essentially persist forever in our simulations except for their interaction with non-axisymmetric spiral density waves excited by a background (Papaloizou-Pringle) instability (which arises in our setup with rigid walls), in addition to numerical viscosity. Presumably the eccentricity would be damped on the disc (turbulent) viscous timescale, but our simulations are not run for this duration (the viscous timescale is very long because the background instability transports angular momentum only weakly). Eccentricity can also be affected by non-adiabatic thermal processes, not considered in this paper. Indeed, \citep{Statler2001} has argued that a uniform eccentricity is preferred because the surface density of the disc is then constant around each orbit. However, even in this case the fluid undergoes periodic compression because of the variation of vertical gravity around the elliptical orbit (which is not captured in two dimensions).

Our observation that gas discs with free eccentricities can remain eccentric for thousands of orbital periods (Fig.~\ref{0}) is potentially important regarding the interpretation of observed large-scale asymmetries in gas discs. In particular, the presence of such asymmetries is often used to infer the presence of perturbing planets (e.g.~\citealt{Regaly2014,Hash2015,Pinilla2015}). However, since we have shown that eccentric modes can be long-lived features, there are alternative mechanisms in addition to perturbing planets that can induce free eccentricity, e.g.~gravitational instabilities in the gas disc during the pre-main sequence phase \citep{Adamsetal1989}.

However, it should be noted that any conclusions regarding the longevity of disc eccentricities are based on two-dimensional simulations with only a weak background instability. In three dimensions eccentric discs are subject to a parametric instability that excites small-scale inertial waves \citep{John2005a,BO2014a}. This is expected to lead to enhanced damping of the disc eccentricity and eccentricity gradients \citep{John2005b}, but further work adopting a realistic vertical disc structure is required to determine its nonlinear outcome. The simulations presented in this paper provide a starting point to allow us to undertake such calculations.

\section*{Acknowledgements}
AJB is supported by the Leverhulme Trust and Isaac Newton Trust through the award of an Early Career Fellowship. The early stages of this research were supported by STFC through grants ST/J001570/1 and ST/L000636/1. We would like to thank the referee for constructive comments that led us to strengthen our conclusions.

\appendix 
\section{Nonlinear secular theory of eccentric discs}
\label{nonlineartheory}
In this section we derive analytically a nonlinear secular theory for eccentric discs in two dimensions utilising the local model of \cite{OB2014a}, for the case of an untwisted isothermal disc in which $\lambda\omega^{\prime}=0$, i.e.~the orbits are aligned (for which nonlinear pressure effects are likely to be minimised). While the formalism can be extended to the case of a twisted disc including viscosity, and to discs with different thermodynamic behaviour \citep{Ogilvie2001}, we do not believe it to be possible to derive the corresponding nonlinear theory analytically in terms of algebraic functions (particularly the stress integrals) except for this special case. We denote $\cos\theta$ by $c$ and $\sin\theta$ by $s$ to simplify the expressions below, where the $\theta=\phi-\omega(\lambda)$, is the true anomaly. At each $\lambda$, the reference orbit is Keplerian, so
\begin{eqnarray}
R = \lambda \left(1+ e c\right)^{-1},
\end{eqnarray}
and the angular velocity of the orbital motion is
\begin{eqnarray}
\Omega=\sqrt{\frac{GM}{\lambda^3}}\left(1+ e c\right)^{2}.
\end{eqnarray}
The orbital period for a fluid element with this orbit is
\begin{eqnarray}
\mathcal{P}=\int_{0}^{2\pi}\Omega^{-1} \mathrm{d}\theta=2\pi\sqrt{\frac{\lambda^{3}}{GM}} (1-e^2)^{-\frac{3}{2}},
\end{eqnarray}
and its specific angular momentum is $\ell=\sqrt{G M\lambda}$.

The evolution of the mass, angular momentum and eccentricity of the eccentric disc are governed by the following stress-integrals, which we evaluate for a general untwisted eccentric disc with non-intersecting orbits:
\begin{eqnarray}
\iint J R^2 T^{\lambda\phi} \mathrm{d} \phi \, \mathrm{d}z &=& 0, \\
\nonumber
\iint J R^2 T^{\lambda\phi} \mathrm{e}^{\mathrm{i}\phi} \mathrm{d} \phi \, \mathrm{d}z &=& \frac{\mathrm{i} c_{s}^2 e \left(e^2-1\right) \lambda \mathcal{M}}{(\lambda e^{\prime})^2}\Bigg( e^2-e \lambda e^{\prime}-1 \\ && \hspace{0.0cm} +\sqrt{\left(e^2-1\right) \left((e-\lambda e^{\prime})^2-1\right)} \Bigg),\\
\nonumber
\iint J R_{\lambda} T^{\lambda\lambda} \mathrm{e}^{\mathrm{i}\phi}\mathrm{d} \phi \, \mathrm{d}z &=& \frac{c_{s}^{2} \left(1-e^2\right)^{3/2} \mathcal{M}}{(\lambda e^{\prime})^2} \Bigg( \frac{\big(e+\lambda e^{\prime}-e^3\big)}{\sqrt{1-e^2}} \\
&& -\frac{\big(e+\lambda e^{\prime}+e^2\lambda e^{\prime} -e^3\big)}{\sqrt{1-\left(e-\lambda e^{\prime}\right)^2}}\Bigg), \\  
\iint J R^2 T^{\phi\phi} \mathrm{e}^{\mathrm{i}\phi}\mathrm{d} \phi \, \mathrm{d}z &=& c_{s}^2 e \mathcal{M},
\end{eqnarray}
where $T^{ij}=-pg^{ij}$ and $J=R R_{\lambda}$. In each case the integrals are over the full extent of the disc in $\phi$ and $z$. We have defined $R_\lambda\equiv \partial_\lambda R$ and $R_\phi\equiv \partial_\phi R$, so that the relevant components of the metric tensor are
\begin{eqnarray}
g^{\lambda \lambda}= \frac{R^2+R_\phi^2}{R^2 R_{\lambda}^2}, \;\; g^{\lambda \phi}= -\frac{R_\phi}{R^2R_\lambda}, \;\; g^{\phi\phi}= \frac{1}{R^2}.
\end{eqnarray}
The one-dimensional mass density is
\begin{eqnarray}
\mathcal{M}=\iint J \Sigma  \, \mathrm{d}\phi.
\end{eqnarray}

The mass and angular momentum do not evolve since the first integral above vanishes, which results from us neglecting shear viscosity. The complex eccentricity evolves according to 
\begin{eqnarray}
\label{nonlinearsecular}
\nonumber
  \ell\mathcal{M}\partial_t E&=&\iint\Big(2\mathrm{e}^{\mathrm{i}\phi}\partial_\lambda(JR^2T^{\lambda\phi})-\mathrm{i}\lambda\,\mathrm{e}^{\mathrm{i}\phi}\partial_\lambda(JR_\lambda T^{\lambda\lambda}) \\ && \hspace{0.4cm}
  -\mathrm{i}\,\mathrm{e}^{\mathrm{i}\phi}JR^2T^{\phi\phi}-\frac{JR^2}{\lambda}\mathrm{e}^{\mathrm{i}\phi}T^{\lambda\phi}\Big)\,\mathrm{d}\phi\,\mathrm{d} z,
\label{edot}
\end{eqnarray}
which is a type of nonlinear Schr\"{o}dinger differential equation (that is second order in $\lambda$). 

For general initial conditions, the time-evolution will produce a twist in the disc, but it is possible to seek modal solutions that remain untwisted. In this case, inputting the stress integrals above leads to an equation for the evolution of the complex eccentricity for any eccentricity and eccentricity gradient for an untwisted disc, as long as the orbits do not intersect. The final form of this equation after substituting the stress integrals is too complicated to be worth writing down in closed form here. It can, however, be computed in Mathematica and exported to text format for input in our Matlab scripts that are used to solve Eq.~\ref{nonlinearsecular}.

We seek modes $E\propto \mathrm{e}^{\mathrm{i}\omega_p t}$ that precess slowly at the rate $\omega_p$, so that Eq.~\ref{nonlinearsecular} is converted to an eigenvalue problem of the form
\begin{eqnarray}
\omega_p e= f(\lambda, e,e',e''),
\end{eqnarray}
subject to appropriate boundary conditions, where $f$ is in general a nonlinear function of its arguments.

The behaviour of the stress integrals for large $e$, $e'$ is straightforward to understand. As the orbits become closer to intersecting ($(e-\lambda e^{\prime})^2\rightarrow 1$) or $e\rightarrow 1$ or $(\lambda e^{\prime})^2\rightarrow 1$, strong pressure forces arise, which modify the eccentricity profile so that the orbits do not intersect. An interesting result of these forces is that they modify the global precession rate of the mode throughout the disc so that it precesses more rapidly in a retrograde sense as the orbits become closer to intersecting, even if this near intersection only occurs in a small region of the disc. We note that the formalism of \cite{BGT1983} for narrow eccentric rings (with small eccentricities but arbitrary eccentricity gradients) applied to an isothermal gas also predicts a precession rate (their Eq.13) that depends on the eccentricity gradient, and which diverges in the limit of intersecting orbits. This is consistent with our results.

The 2D isothermal model is a simplification of a realistic 3D eccentric disc. The isothermal case (with adiabatic index of 1) is the most compressible 2D model that we can consider, where the pressure increase for a given streamline convergence is minimised (over less compressible models with adiabatic indices larger than 1). However, in 3D the disc can also expand vertically, which may somewhat alleviate this pressure increase. However the qualitative behaviour of the 2D isothermal model is likely to carry over to more realistic models.

\section{Linear non-secular theory in 2D}
\label{nonseculartheory}
As listed in Table~\ref{table1}, we solve the eigenvalue problem for the full (non-secular) isothermal hydrodynamical equations in two dimensions to compute the precession frequency of the fundamental $m=1$ linear eccentric mode, to allow comparison with simulations and secular theory. We assume linear perturbations $u_R^{\prime},u_{\phi}^{\prime},\Sigma^{\prime}\propto \mathrm{e}^{\mathrm{i}(\phi-\omega_p t)}$, which satisfy
\begin{eqnarray}
&&-\mathrm{i}\hat{\omega}u_R^\prime =2\Omega u_{\phi}^{\prime}-\frac{c_s^2}{\Sigma_b}\partial_R \Sigma^{\prime}+\frac{c_s^2 \Sigma^{\prime}}{\Sigma_b^2}\partial_R\Sigma_b, \\
&&-\mathrm{i}\hat{\omega}u_\phi^\prime=-\left(2\Omega+R\partial_R\Omega\right)u_R^{\prime}-\frac{\mathrm{i}c_s^2}{R}\Sigma^{\prime},\\
&&-\mathrm{i}\hat{\omega}\Sigma^{\prime}=-\Sigma_b \left(\frac{u_R^{\prime}}{R}+\partial_R u_{R}^{\prime}\right)-u_R^{\prime}\partial_R\Sigma_b - \frac{\mathrm{i}\Sigma_b}{R}u_{\phi}^{\prime},
\end{eqnarray}
where $\hat{\omega}=\omega_p-\Omega$, subject to the BCs that $u_R^{\prime}=0$ at $R=R_i$ and $R=R_o$. In this case 
\begin{eqnarray}
\label{Omega}
\Omega &=& \Omega_{0} R^{-\frac{3}{2}}\left(1-\frac{\sigma c_{s}^{2}}{\Omega^2_0}R\right)^{\frac{1}{2}}, 
\end{eqnarray}
where we have taken into account the radial pressure gradient in full. This is solved using a Chebyshev collocation method (adopting 101 points in radius, which we find to be more than sufficient to compute the fundamental mode), which converts the problem to a standard eigenvalue problem that can be solved (for which we use a QZ method) to compute the precession frequency $\omega_p$.

\bibliography{disc}
\bibliographystyle{mn2e}
\label{lastpage}
\end{document}